\documentclass[fleqn,usenatbib]{mnras}

\usepackage{newtxtext,newtxmath}

\usepackage[T1]{fontenc}

\DeclareRobustCommand{\VAN}[3]{#2}
\let\VANthebibliography\thebibliography
\def\thebibliography{\DeclareRobustCommand{\VAN}[3]{##3}\VANthebibliography}

\usepackage{graphicx}	
\usepackage[normalem]{ulem}

\usepackage{tabularx}
\usepackage{multirow}
\usepackage{xcolor}

\usepackage{graphicx}	
\usepackage{amsmath}	

\newcommand{\krot}{$\kappa_\mathrm{rot}$}
\newcommand{\hMpc}{$h^{-1}$Mpc}
\newcommand{\Msol}{M$_{\odot}$}
\newcommand{\DIA}{$D_{\rm{IA}}$}

\renewcommand{\vec}[1]{ {\bf #1} }

\title[MTNG Intrinsic Alignments]{The MillenniumTNG Project: Intrinsic alignments of galaxies and halos}

\author[A.M. Delgado et al.]{Ana Maria Delgado,$^{1}$\thanks{E-mail: ana\_maria.delgado@cfa.harvard.edu (AMD)}
Boryana Hadzhiyska$^{1,2,3}$,
Sownak Bose$^{4}$,
Volker Springel$^{5}$,
Lars Hernquist$^{1}$,
\newauthor%
Monica Barrera$^{5}$,
R\"udiger Pakmor$^{5}$,
Fulvio Ferlito$^{5}$,
Rahul Kannan$^{1,6}$,
C\'esar Hern\'andez-Aguayo$^{5,7}$,
\newauthor%
Simon D. M. White$^{5}$,
and Carlos Frenk$^{4}$
\\%
\\%
$^{1}$Harvard-Smithsonian Center for Astrophysics, 60 Garden St, Cambridge, MA 02138, USA\\%
$^{2}$Miller Institute for Basic Research in Science, University of California, Berkeley, CA, 94720, USA\\%
$^{3}$Physics Division, Lawrence Berkeley National Laboratory, Berkeley, CA 94720\\%
$^{4}$Institute for Computational Cosmology, Department of Physics, Durham University, South Road, Durham, DH1 3LE, UK\\%
$^{5}$Max-Planck-Institut f\"{u}r Astrophysik, Karl-Schwarzschild-Str. 1, 85748, Garching, Germany\\%
$^{6}$Department of Physics and Astronomy, York University, 4700 Keele St., Toronto, ON M3J 1P3, Canada \\%
$^{7}$Excellence Cluster ORIGINS, Boltzmannstrasse 2, 85748 Garching, Germany%
}
\date{Accepted XXX. Received YYY; in original form ZZZ}

\pubyear{2021}

\begin{document}
\label{firstpage}
\pagerange{\pageref{firstpage}--\pageref{lastpage}}
\maketitle

\begin{abstract}
The intrinsic alignment (IA) of observed galaxy shapes with the underlying cosmic web is a source of contamination in weak lensing surveys. Sensitive methods to identify the IA signal will therefore need to be included in the upcoming weak lensing analysis pipelines. Hydrodynamical cosmological simulations allow us to directly measure the intrinsic ellipticities of galaxies and thus provide a  powerful approach to predict and understand the IA signal. Here we employ the novel, large-volume hydrodynamical simulation MTNG740, a product of the MillenniumTNG (MTNG) project, to study the IA of galaxies. We measure the projected correlation functions between the intrinsic shape/shear of galaxies and various tracers of large-scale structure, $w_{+g},\ w_{+m},\ w_{++}$ over the radial range $r_{\rm p} \in [0.02 , 200]\,h^{-1}{\rm Mpc}$ and at redshifts $z=0.0$, $0.5$ and $1.0$. We detect significant signal-to-noise IA signals with the density field for both elliptical and spiral galaxies. We also find significant intrinsic shear-shear correlations for ellipticals. We further examine correlations of the intrinsic shape of galaxies with the local tidal field. Here we find a significant IA signal for elliptical galaxies assuming a linear model. We also detect a weak IA signal for spiral galaxies under a quadratic tidal torquing model. Lastly, we measure the alignment between central galaxies and their host dark-matter halos, finding small to moderate misalignments between their principal axes that decline with halo mass.

\end{abstract}




\section{Introduction}
\label{Intro}
Weak gravitational lensing has been recognized as an important probe of the cosmic large-scale structure \citep{2001PhR...340..291B, 2007MNRAS.381.1018A, 2015RPPh...78h6901K}. Its sensitivity to both the luminous and dark matter components of our Universe \citep{2004ApJ...600...17B, 2004ApJ...601L...1T} make it an effective tool with which to not only investigate the nature of dark matter and dark energy, providing a way to constrain the cosmological parameters \citep{2012PhRvD..85j3518V}, but also for the investigation of modified theories of gravity \citep{2008JCAP...04..013A,2011PhRvD..83b3012M} and for obtaining independent constraints on the growth and expansion history of the Universe \citep{2017PhRvD..96b3513Z}. Recent weak lensing surveys, such as KiDS \citep{2018MNRAS.474.4894J, 2020A&A...638L...1J} and DES \citep{2018PhRvD..98d3526A}, have provided constraints on the $\Lambda$CDM paradigm that are in relatively good agreement with observations from the Cosmic Microwave Background (CMB). However, tensions, such as in the constraints for $\Omega_{\rm{m}}$ and $\sigma_8$, still remain \citep{2015MNRAS.451.2877M, 2019arXiv190105289D}. 

The next generation of weak lensing surveys, such as Euclid \citep{2011arXiv1110.3193L}, the Vera~Rubin Observatory \citep{2009arXiv0912.0201L} and the Nancy Roman Space Telescope \citep{2015arXiv150303757S}, will provide us with measurements of galaxy shapes for unprecedentedly large samples. The main limitation to our interpretation of the data will thus come, not from a lack of statistical information, but from systematic errors in our analysis pipelines. The intrinsic alignments of galaxies with the surrounding matter distribution is one such systematic that needs to be considered in weak lensing analyses. Systematic shape distortions of galaxies in a given field of view can be present at a certain level even in the absence of lensing. If unaccounted for, this intrinsic alignment can adversely affect the utility of weak lensing as a tool for precision cosmology and lead to biases in inferences \citep{PhysRevD.70.063526, PhysRevD.82.049901,2007NJPh....9..444B, 2012MNRAS.424.1647K,2015RPPh...78h6901K}. In order to fully realize the potential of weak lensing as a cosmological probe, intrinsic alignment signals must therefore be accounted for and marginalized over \citep{2010A&A...523A...1J, 2010MNRAS.406L..95Z, 2012MNRAS.419.1804T}. Furthermore, a number of works have pointed out that there is cosmological information to be extracted from the intrinsic alignment signal itself \citep{2013JCAP...12..029C, 2014PhRvD..90d3527C, 2015JCAP...10..032S}. 

Galaxy formation and evolution processes are linked at various stages to the creation and potential destruction of galaxy alignments, but the factors determining the intrinsic alignment signals are not fully understood. Some early studies focused on the dependence of alignment signals on galaxy morphology and used $N$-body simulations to identify galaxy types and measure intrinsic alignment signals. For example, \cite{2000MNRAS.319..649H} assumed that the orientation of a disk galaxy is set by the angular momentum of its host halo, and focused on measuring a correspondingly expected alignment signal from spiral galaxies. Meanwhile, \citet{2000ApJ...545..561C} assumed that elliptical galaxies inherit their shape from their host halos and concentrated on detecting the intrinsic alignment signal from elliptical galaxies. 

In recent years, hydrodynamical simulations have become a key tool for studying intrinsic alignments. They are advantageous because their galaxy formation physics models have been able to predict increasingly realistic galaxies without relying on assumptions about the occupation of halos with galaxies and about their alignments with the host halo. Furthermore, hydrodynamical simulations can be analyzed free of lensing effects; therefore, all alignments measured are characteristically ``intrinsic'' and unaffected by weak lensing. They are therefore particularly well-suited for studying the dependence of intrinsic alignments on various properties that may affect the observed shape of galaxies. \citet{2015MNRAS.454.2736C} investigated intrinsic alignments at $z=0.5$ using the Horizon-AGN \citep{2014MNRAS.444.1453D} simulation by way of projected shape correlations and found dependencies on galaxy shape based on colour, as well as a mass dependence in the alignment of ellipticals with central galaxies. \citet{2015MNRAS.448.3522T} similarly measured intrinsic alignments and found mass and redshift dependencies using MassiveBlack-II \citep{2015MNRAS.450.1349K}, over the range $z=0.0-1.0$. \citet{2017MNRAS.468..790H} presented a comprehensive analysis of intrinsic alignments in the Illustris \citep{2014Natur.509..177V} simulation, finding stronger correlations for more massive and more luminous galaxies, as well as for earlier photometric types, in agreement with observations. \cite{2021JCAP...03..030S} found dependencies in on redshift, galaxy mass, environment and morphology usint IllustrisTNG. \cite{2021MNRAS.501.5859T} further detected increased projected correlations for satellite galaxies tracing more massive central galaxies, as well as a colour dependence consistent across MassiveBlack-II and IllustrisTNG \citep{2018MNRAS.475..624N} at low redshifts. The central galaxy mass dependence of satellite alignment was also found in \cite{2015MNRAS.454.3328V} using EAGLE \citep{2015JCAP...10..032S} and cosmo-OWLS \citep{2014MNRAS.441.1270L} at $z=0.0$. While \cite{2022MNRAS.511.3844H} used EAGLE to study the intrinsic alignments of the star-forming gas component of galaxies and found that it exhibits weaker intrinsic alignments in three dimensions than do the stellar and dark-matter components of galaxies.

Recent work by \citet[][henceforth Z22]{2022MNRAS.514.2049Z} used TNG100 to examine the correlations of galaxy shapes and local tidal shear tensor and to test analytic intrinsic alignment mechanisms for elliptical and spiral galaxies at $z=0.0$ and $z=1.0$. They found a significant alignment strength for elliptical galaxies under a linear alignment model at both redshifts, and for spiral galaxies at $z=1.0$. They did not, however, detect any intrinsic alignment for spiral galaxies under a non-linear model. They further reported dependencies of the intrinsic alignment strength on various properties such as redshift, mass and smoothing scale of the tidal shear tensor. 

In a recent comparison study, \citet{2021MNRAS.508..637S} measured and compared projected intrinsic shape correlations in the Illustris, MassiveBlack-II, and TNG300 simulations, reporting non-zero intrinsic alignment at the level of several sigma for all three simulations, but with considerable noise and potentially systematic differences among the simulations. They also did not find evidence for a non-zero tidal torquing amplitude for TNG300, but did marginally detect such a signal for MB-II.

In this work, we make use of the substantially increased statistical power of the new MillenniumTNG (MTNG) hydrodynamical simulation, which improves the volume by a factor of $\sim$300  compared to TNG100 and a factor of $\sim$15 compared to TNG300. This provides an opportunity to study intrinsic alignment signals directly from cosmological galaxy formation simulations out to much larger projected distances than possible thus far. In addition, we also provide, for the first time, simulation-based measurements of the intrinsic shape correlations expected between both elliptical and spiral galaxies and the full non-linear matter density field. This should be of help in testing and refining analytic models for the origin of intrinsic alignment signals. In this vein, we also expand upon the study of Z22 and consider to what extent our hydrodynamical simulation exhibits galaxy shape correlations with the local tidal field, something that is often assumed in analytic theories for intrinsic alignment \citep[e.g][]{PhysRevD.70.063526, 2009IJMPD..18..173S, 2012MNRAS.421.2751S}.

The organization of this paper is as follows. In Section~\ref{Methods}, we introduce the flagship MillenniumTNG hydrodynamical simulation, as well as our methods for galaxy selection, our definition of galaxy ellipticity, our measurement approach for projected intrinsic shear correlations, and our definitions of parameters for measuring intrinsic alignment with the local tidal tensor. In Section~\ref{results:correlations} we present our results on projected correlation functions of the intrinsic shapes of galaxies. We then give in Section~\ref{IA_gals} our results on the intrinsic alignment of both elliptical and spiral galaxies, and we examine the dependence of intrinsic alignment strength on galaxy and environmental properties. In Section~\ref{IA_halos} we consider the intrinsic alignment of dark-matter halos, and in Section~\ref{Misalignmnet} we inspect the misalignment between host halos and their central galaxies. We summarize and conclude in Section~\ref{conclusions}.

\section{Methods}
\label{Methods}
\subsection{The MilleniumTNG simulations}
\label{Methods:MTNG}

We employ the flagship, full physics simulation from the newly created suite of hydrodynamical and cosmological dark-matter-only simulations of the MilleniumTNG project. This full physics simulation, henceforth MTNG740, is a periodic box with comoving volume (500 \hMpc~$\approx 740\, \mathrm{Mpc}$)$^3$ using $4320^3$ dark-matter particles with mass $1.7\times10^8\,$\Msol, and $4320^3$ gas cells, each with an initial mass of $3.1\times10^7\,$\Msol. The physics model of MTNG740 is based on the IllustrisTNG \citep{2018MNRAS.475..648P, 2018MNRAS.475..676S, 2018MNRAS.480.5113M, 2018MNRAS.475..624N, 2018MNRAS.477.1206N} galaxy formation physics model \citep{2017MNRAS.465.3291W, 2018MNRAS.473.4077P, 2019MNRAS.490.3196P, 2019ComAC...6....2N,2019MNRAS.490.3234N} which has been shown to produce realistic galaxy populations on cosmological scales. Furthermore, all cosmological parameters are given by \cite{2016A&A...594A..13P} as in IllustrisTNG: $\Omega_0=0.3089$ (total matter density), $\Omega_{\rm b}=0.0489$ (baryon density), $\Omega_{\Lambda}=0.6911$ (dark-energy density), $H_0=67.74\,{\rm km\, s^{-1}Mpc^{-1}}$ (Hubble parameter) and $\sigma_8=0.8159$ (linear rms density fluctuations in a sphere of radius 8 \hMpc~ at $z=0.0$). Initial conditions were generated at $z=63$ with second-order Lagrangian perturbation theory using {\small GADGET-4} \citep{2021MNRAS.506.2871S}. All full-physics simulations in the suite were run using the {\small AREPO} code \citep{2010MNRAS.401..791S,2020ApJS..248...32W}. We note that in order to reduce the memory requirements of the MTNG740 simulation, minor modifications were made to the physics model as compared to IllustrisTNG; namely, magnetic fields and a tracing of individual metal species needed to be disabled in MillenniumTNG. These changes, particularly removing the magnetic fields, have a slight but measurable impact on the simulation results for galaxy properties \citep{2017MNRAS.469.3185P}.

Many typical post-processing tasks were performed on-the-fly while running the MTNG suite, which facilitates working with the very large MilleniumTNG simulations. This includes halo finding using the Friends-of-Friends (FoF) algorithm and identifying gravitationally bound substructures using the new {\small SUBFIND-HBT} algorithm \citep{2021MNRAS.506.2871S}. Galaxies are defined as bound systems of stars based on the structures identified by {\small SUBFIND-HBT}. Several other data products that were generated on-the-fly include power spectra, merger trees and a series of lightcones in a variety of geometries. For further details on the technical specifications of the MillenniumTNG suite and its outputs we refer the reader to \cite{2022arXiv221010059H} and \cite{2022arXiv221010060P}.

\begin{figure*}
    \centering
    \includegraphics[width=0.95\textwidth]{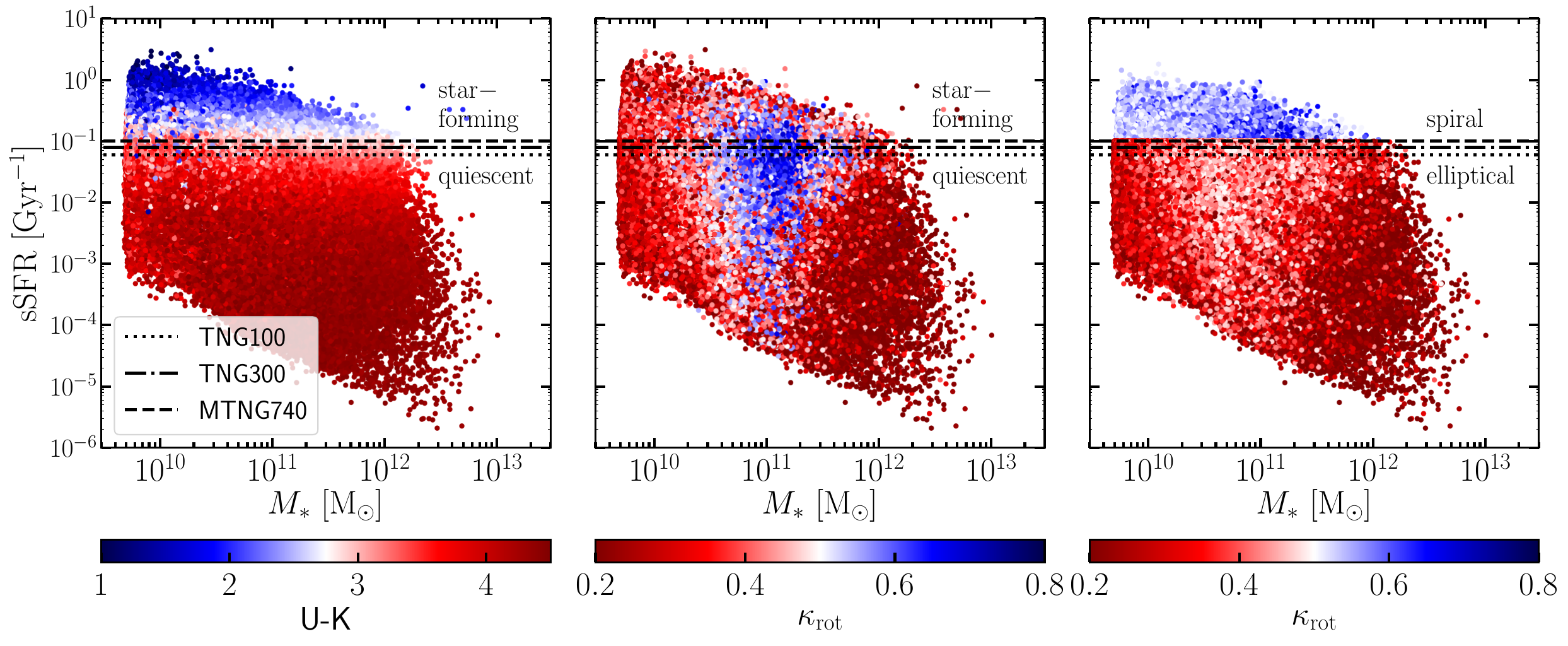}
    \caption{Galaxy selection following the Z22 prescription. We show three plots with identical axes, sSFR along the $y$-axis and stellar mass along the $x$-axis. We define our elliptical galaxies in two steps. We first make a cut based on colour as a proxy for star-forming or quiescent galaxies (left). We show the sSFR cut-off for quiescent, red galaxies for MTNG to be at sSFR $\sim$ 0.1 Gyr$^{-1}$. We then consider the amount of rotational support of the galaxy as given by the \krot~parameter (center). We define elliptical galaxies as those having $\kappa_{\rm rot} \leq 0.5$. The right plot shows our final selection of elliptical galaxies in red, based both on sSFR and \krot. }
    \label{fig:galaxy_selection}
\end{figure*}

The work presented here is part of a series of introductory papers associated with the MillenniumTNG project. \cite{2022arXiv221010059H} provide a detailed introduction to the project with an analysis of matter clustering and halo statistics, while \cite{2022arXiv221010060P} present a detailed introduction to the full physics simulations with a focus on the scaling relations of large clusters. In \cite{2022arXiv221010419B} we introduce a new method of producing semi-analytic galaxy formation predictions for the lightcone outputs. \cite{2022arXiv221010065B} give an analysis of galaxy clustering while \cite{2022arXiv221010075C} use the MTNG simulations to test methods for inferring cosmological parameters from galaxy clustering. \cite{2022arXiv221010068H, 2022arXiv221010072H} present an analysis of, and improvement to, HOD techniques with respect to galaxy assembly bias for a selection of DESI-like galaxies in MTNG. \cite{2022arXiv221010066K} provide an analysis of high redshift galaxies and compare the results to recent observations made by the James Webb Space Telescope. Finally, in \cite{Ferlito2022} we examine the weak gravitational lensing convergence maps both in full physics and dark-matter-only runs.

\subsection{Galaxy selection}
\label{Methods:Galaxies}

This work focuses on galaxy populations defined by their luminosity and morphology, similar to those being targeted by upcoming surveys. Using the MTNG740 subhalo catalog, we define our galaxies as subhalos with a minimum total mass of $1.0 \times 10^{10}\, \mathrm{M}_{\odot}$ and at least 300 stellar particles, giving a minimum stellar mass of $\sim 1.55 \times 10^{9}\, \mathrm{M}_{\odot}$. To identify galaxy morphology, we use a similar prescription to Z22: elliptical galaxies are identified as quiescent, red galaxies whose stability is provided by the velocity dispersion of stellar particles, whereas spirals are defined as star-forming, blue galaxies with stellar distributions supported primarily by rotation. 

Fig. \ref{fig:galaxy_selection} depicts how we identify our elliptical and spiral galaxy populations. In each panel, we show the specific star formation rate (sSFR) as a function of stellar mass for each galaxy in MTNG740. The left panel is colour-coded by the U-K colour index in order to distinguish between blue and red galaxies. Quiescent galaxies are thus defined as those with ${\rm sSFR} < 0.1\,{\rm  Gyr}^{-1}$ (the red galaxies), and star-forming galaxies are those with ${\rm sSFR} > 0.1\,{\rm  Gyr}^{-1}$ (blue galaxies). We include horizontal lines showing the sSFR cut-off for red vs.~blue galaxies in TNG100 and TNG300 for reference, and note the similarity in their values of sSFR compared to MTNG740 despite the differences in resolution across the three simulations. We see that as galaxy mass increases the sSFR decreases, resulting in two main populations: a population of high stellar mass, quenched galaxies, and another of low-mass, star-forming galaxies.

The middle panel of Fig.~\ref{fig:galaxy_selection} is colour-coded by the amount of rotational support defined by the dimensionless parameter
\begin{equation}
    \kappa_\mathrm{rot} = \frac{E_{\rm rot}}{E_{\rm kin}} = \sum_n \frac{1}{2}m_n\left(\frac{j_{z,n}}{r_n} \right)^2 / \sum_n \frac{1}{2}m_n\textbf{v}_n^2,
\end{equation}
where $E_\mathrm{rot}$ and $E_\mathrm{kin}$ are the rotational and kinetic energy of the stellar galaxy, respectively, $m_n$ denotes the mass of stellar particle $n$, $j_{z,n}$ is its specific angular momentum in the $z$-direction (defined as the minor axis of the galaxy's shape), $r_n$ is its distance to the center of the galaxy, and $\textbf{v}_n$ is its velocity in the center-of-mass frame. Here, we find that the galaxies with the least rotational support, according to the value of \krot, tend to fall toward the extremes of the mass range (low mass and high mass), while galaxies that have the highest rotational support fall in the intermediate range of stellar masses $\sim 10^{11} \mathrm{M}_{\odot}$. This is consistent with Z22 and other work \citep{2015MNRAS.454.2736C, 2017MNRAS.467.3083R}.

The right panel in Fig.~\ref{fig:galaxy_selection} shows our final selection of spiral and elliptical galaxies. Spiral galaxies are defined as those with ${\rm sSFR}>0.1\, {\rm Gyr}^{-1}$ and $\kappa_{\rm rot}> 0.5$, while elliptical galaxies are those with ${\rm sSFR}<0.1\, {\rm Gyr}^{-1}$ and $\kappa_{\rm rot}< 0.5$. We conservatively exclude galaxies whose $\rm{sSFR}=0.1$ and $\kappa_{\rm rot}=0.5$ from  either population when performing studies on the dependence of galaxy morphology. This leaves us with sample sizes of $\sim$541,000 ellipticals and $\sim$145,000 spirals at $z=0.0$, $\sim$276,000 ellipticals and $\sim$196,000 spirals at $z=0.5$, and $\sim$139,000 ellipticals and $\sim$182,000 spirals at $z=1.0$. For studies using all galaxies combined, however, we do not make any cuts based on $\rm{sSFR}$ or $\kappa_{\rm rot}$ and obtain sample sizes of $\sim 1.3\times10^6$, $\sim1.2\times10^6$ and $\sim9.9\times10^5$  galaxies in total at $z=0.0,\ z=0.5$ and $z=1.0$, respectively.

\subsection{Projected galaxy ellipticity}
\label{Methods:ellipticity}

We observe the shapes of galaxies through the brightness of their stars. We therefore determine a measure for 3D galaxy shape by using luminosity-weighted stellar positions in the V-band. The second moment tensor of the brightness distribution can be defined as \citep{1983ApJ...271..431V, 1991ApJ...380....1M}:
\begin{equation}
    q_{ij}^{(b)}=\frac{1}{L^{(b)}}\sum_n \left(\vec{x}_n-\overline{\vec{x}}^{(b)}\right)_i \left(\vec{x}_n-\overline{\vec{x}}^{(b)}\right)_jL_n^{(b)}
    \label{equ.brightness_tensor} ,
\end{equation}
where $L^{(b)}$ is the total luminosity of the galaxy in the photometric band $b$ (we adopt the V-band in this work), $\vec{x}_n$ is the position vector of stellar particle $n$, $\overline{\vec{x}}^{(b)}$ is the luminosity weighted center of the galaxy in band $b$, and $L_n$ is the luminosity of the stellar particle in this band. 

From these 3D moments of the brightness distribution, one can also readily determine projected moments of the light distribution, for example along the  Cartesian $z$-axis of the simulation box, which we in the following adopt as our default projection direction\footnote{We will repeat our analysis also for projections along the $x$-axis and $y$-axis, and use the variations among them as a rough measure of statistical uncertainty.}. In particular, we define the projected, complex ellipticity  $\epsilon$ of the galaxy along the Cartesian $z$-axis, pertinent in weak lensing analysis, as follows:
\begin{equation}
    \epsilon = \epsilon_{+} + {\rm i} \epsilon_{\times} =  \frac{q_{xx}-q_{yy}}{q_{xx}+q_{yy}} + {\rm i}\frac{2q_{xy}}{q_{xx}+q_{yy}}
    \label{eq.ellpiticity},
\end{equation}
where we have omitted the $(b)$ for notational simplicity. $\epsilon_{+}$ refers to the elongation along the coordinate system, while $\epsilon_{\times}$ measures the elongation at a 45 degree angle from the axes. $\epsilon$ transforms under rotation as a traceless, symmetric $2\times 2$ tensor, i.e.~for a rotation under an angle $\theta$, $\epsilon$ changes as $\epsilon'= \epsilon \exp(2 {\rm i}\theta)$. The ellipticity is thus invariant under a rotation by 180 degrees, as expected, and changes sign under a rotation by 90 degrees.

\begin{figure}
    \centering
    \includegraphics[width=0.99\columnwidth]{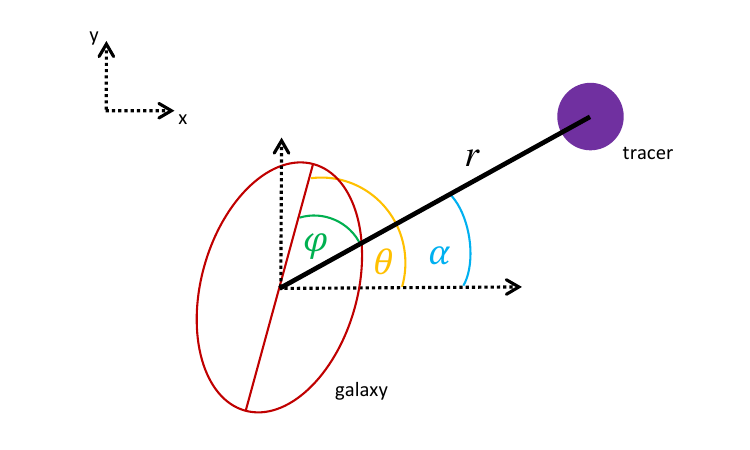}
    \caption{Illustration of the alignment angle $\varphi$. Tracer and galaxy are separated by a distance $r$, where the angle of the separation vector with respect to the $x$-direction is given by $\alpha$. The angle of the galaxy orientation, with respect to the $x$-direction is given by $\theta$. The orientation angle of the galaxy with respect to the separation vector is thus given by $\varphi$ = $\theta - \alpha$.}
    \label{fig:IA_cartoon}
\end{figure}

\subsection{Correlations of intrinsic shear}

Two-point correlations of galaxy shapes are one of the main probes of gravitational weak lensing surveys. However, if the orientations of the intrinsic ellipticities of galaxies correlate with the surrounding distribution of matter even in the absence of lensing, such intrinsic alignments can systematically bias cosmological inferences from weak lensing. The most straightforward approach to quantify such intrinsic alignments is therefore to measure the projected two-point correlation function of the intrinsic ellipticity of galaxies with the surrounding density field (this can either be traced by the same or a different population of galaxies, or it can be the actual matter field itself). Even more directly one can consider the correlations of the ellipticities of galaxies with the ellipticities of their neighbours. 

In weak lensing analysis, it is often assumed that the convergence $\kappa$ and the lensing shear $\gamma$ are weak enough that the reduced shear $g=\gamma/(1-\kappa)$ can be simply approximated by the shear $\gamma$. Since in this limit the observed ellipticity effectively arises as a sum of the intrinsic ellipticity and the lensing-induced ellipticity \citep{2001PhR...340..291B, 2017MNRAS.468..790H}, it is customary to convert the intrinsic ellipticity into an intrinsic shear, through
\begin{equation}
\gamma = \frac{\epsilon}{2 {\cal R}}, 
\end{equation}
in order to more easily compare to the weak-lensing shear. We will thus in the following report intrinsic shear correlation functions based on this conversion\footnote{Alternatively, a different definition of ellipticity \citep[referred to as $\eta$ in][]{2017MNRAS.468..790H} that carries the same information could have been used.}. The responsitivity factor ${\cal R}$ can be used to calibrate the average response of measured galaxy shapes to a given shear \citep{2002AJ....123..583B}, but if the galaxy isophotes are elliptical and no measurement noise is present, ${\cal R}\simeq 1$. We will thus assume ${\cal R} = 1$ in the following. We remind the reader that the intrinsic ellipicity, $\epsilon$, may be rotated under an angle $\theta$ as described in the previous subsection.

Suppose we have a 3D sample of galaxies with complex intrinsic shear $\gamma^{j} = \gamma^{j}_{+} + {\rm i}\, \gamma^{j}_{\times} = \epsilon^{j}/2$ at coordinates $\vec{r}^{j}$, and with mean number density $\overline{n}_{\gamma}$.  Let us further consider a distribution of tracers $n_{\rm T}(\vec{r}) = \sum_k \delta(\vec{r} - \vec{r}_k)$ at coordinates $\vec{r}_k$ and with mean density $\overline{n}_{\rm T}$. The tracer density contrast can then be computed as $\delta_{\rm T}(\vec{r}) = n_{\rm T}(\vec{r}) 
 / \overline{n}_{\rm T} - 1$.

For our galaxies with shapes, let us now define a dimensionless intrinsic shear field relative to a unit direction vector $\vec{\hat{d}}$ in the $xy$-plane as
\begin{equation}
\gamma(\vec{r}, \vec{\hat{d}}) = \frac{1}{\overline{n}_{\gamma}}\sum_j \tilde{\gamma}^j_{+} \,\delta(\vec{r} - \vec{r}_i),
\end{equation}
where $\tilde{\gamma}^j_{+} = {\rm Re} \left\{ \gamma^{j} \exp(-2\, {\rm i}\,\alpha_{\vec{\hat{d}}}) \right\}$ is the ellipticity $\gamma_{+}^j$  of the galaxy expressed relative to the direction defined by vector  $\vec{\hat{d}}$. See Fig. \ref{fig:IA_cartoon} for a sketch of the corresponding geometry. The two-point cross-correlation of the shear field with the tracer field is then defined as 
\begin{equation}
\xi^{+{\rm T}}(r_{\rm p}, \Pi) = \left< \gamma(\vec{r}', \vec{\hat{d}}) \cdot  \delta_{\rm T}(\vec{r}' + r_{\rm p} \vec{\hat{d}} + \Pi\, \vec{\hat{z}})\right>,
\label{eqnxip}
\end{equation}
where $\vec{\hat{z}}$ is the unit vector in the line-of-sight direction, and the expectation value averages over all coordinates $\vec{r}'$  and directions of $\vec{\hat{d}}$. The 3D separation $r$ between a shape and a density field tracer is decomposed here into a projected separation, $r_{\rm p}$, and a line-of-sight separation, $\Pi$, in the usual way, with $r = (r^2_{\rm p} + \Pi^2)^{1/2}$. Note that radial orientations, where the major axes of elliptical shapes tend to point towards tracer particles, are characterized by positive correlations ($\xi^{+{\rm T}} > 0$), while tangential orientations give rise to negative correlations ($\xi^{+{\rm T}} <0$)\footnote{While this sign convention is customary in the IA literature, there are occasionally exceptions \citep[e.g.][]{2015MNRAS.454.2736C} that use the opposite sign convention.}.

Similarly, we can also define shear-shear correlations of the shape sample with itself through
\begin{equation}
\xi^{++}(r_{\rm p}, \Pi) = \left< \gamma(\vec{r}', \vec{\hat{d}}) \cdot  \gamma(\vec{r}' + r_{\rm p} \vec{\hat{d}} + \Pi\, \vec{\hat{z}}, \vec{\hat{d}})\right>.
\end{equation}
The two-point correlations can furthermore be converted into the \textit{projected} correlations, $w_{\rm p}$, by integrating along the line-of-sight, i.e.
\begin{equation}
w^{+{\rm T}}_{\rm p}(r_{\rm p})= \int_{-\Pi_{\rm{max}}}^{\Pi_{\rm{max}}} \xi^{+{\rm T}}(r_{\rm p}, \Pi)\,{\rm d}\Pi,
\label{eqnproj}
\end{equation}
and equivalently for $w^{++}_{\rm p}(r_{\rm p})$ as a projection of $\xi^{++}(r_{\rm p}, \Pi)$. This addresses two practical problems by largely eliminating redshift space distortions in observational data and by removing the inconvenient anisotropy in the 3D quantity $\xi^{+{\rm T}}(r_{\rm p}, \Pi)$. The integration limit $\Pi_{\rm max}$ needs to be sufficiently large such that no contributions to the correlations are dropped by cutting off the integration at some finite distance. We use $\Pi_{\rm{max}}=100\,h^{-1}{\rm Mpc}$ in this work, and have verified that this conservative choice has no influence on our results.

To estimate the value of $w_{\rm p}^{+{\rm T}}$ for a radial bin of a certain finite width $\Delta \ln r_{\rm p}$, we insert the definition of the correlation function into equation~(\ref{eqnproj}) and integrate over the bin in radius, assuming that $w_{\rm p}^{+{\rm T}}$ can be approximated as constant over the size of the bin. Together with the integrations needed to evaluate the expectation value in equation~(\ref{eqnxip}) this effectively yields a volume integration over $\vec{r}'$, and a cylindrical integration over the radial bin and the $z$-range  $[-\Pi_{\rm{max}}, \Pi_{\rm{max}}]$, transforming the products of Dirac $\delta$-functions to simple pair counts. For example, the estimated value of  $w^{+{\rm T}}$ for a certain radial can then be estimated as
\begin{equation}
\left.w^{+{\rm T}}_{\rm p}\right|_{\rm bin} \simeq \frac{\sum_{\rm pairs} \tilde{\gamma}^j_{+}}{ \pi (r_{\rm b}^2 - r_{\rm a}^2)\,\overline{n}_{\gamma} \overline{n}_{\rm T} V},
\end{equation}
where $r_{\rm b}$ and $r_{\rm a}$ are the outer and inner radii of the bin, $V$ is the box volume, and the sum extends over all $j-{\rm T}$ pairs with $xy$-projected separation vector falling into this radial range, and where the corresponding $|\Delta z|$-distance is at most $\Pi_{\rm max}$. The $\tilde{\gamma}_{+}^j$ values are computed individually for each pair using its corresponding direction vector in the $xy$-plane. Note that since we have a well-defined periodic simulation box with uniform sampling density there is no need to introduce a random point set as a helper to estimate the correlation function \citep{1993ApJ...412...64L, 2011MNRAS.410..844M}. In fact, this would only degrade the measurement result in our case. 

In practice, however, carrying out the pair count with a simple double sum over all possible pairs will be computationally extremely expensive, due to the $N^2$-scaling involved. One common approach to reduce the computational cost is to restrict the measurement to relative small maximum distances $r_{\rm p}$, such that only the nearby neighbours need to be considered, which can for example be found through a search grid. However, we do not want to impose such a restriction here,  as this would preclude an accurate measurement of $w^{+{\rm T}}_{\rm p}$ for the particularly interesting regime at large distances, which  is very relevant for weak lensing studies. Note also that we are uniquely positioned to address this large distance regime for the first time directly with a hydrodynamical simulation of galaxy formation, thanks to our unprecedentedly large box size.  Furthermore, we would also like to use the full density field of the simulation as a tracer sample unaffected by galaxy biasing. But the density field of the simulation is sampled by 160 billion resolution elements -- definitely not something suitable for a direct pair count for $\sim$1 million galaxies.

We address this problem by repurposing the hierarchical tree algorithm normally used by the {\small GADGET-4} code \citep{2021MNRAS.506.2871S} to compute the gravitational field. By arranging the tracers into an oct-tree, we can do the pair count by walking the tree. As soon as a tree node falls entirely into one of our logarithmically spaced radial bins, all tracers associated with the corresponding node can be counted in one step, while otherwise the node is opened and its smaller daughter nodes are considered in turn. In this way, the cost of the correlation function measurement effectively reduces to $N_g\log N_{\rm T}$, which is not a problem any more even for an arbitrarily large tracer set $N_{\rm T}$. In addition, if the number of galaxies, $N_{g}$, is very large (say well over $10^6$) one can consider a random subsample of this set as a speed-up measure without significantly increasing the statistical uncertainty of the measurement.

\subsection{Tidal tensor}
\label{Methods:tensor}

It is generally thought that intrinsic alignments of galaxies are sourced by tidal interactions with the gravitational field of larger structures, although additional effects could in principle play a role. On large scales (in the two-halo regime) galaxies may be tidally aligned with the linearly evolving cosmic large-scale structure, while on small scales (the one-halo regime) galaxies may align themselves instead with their host halo. Another common idea is that the alignment of spiral galaxies is related to the angular momentum imprinted on halos by large-scale tidal torquing. 

To examine whether MTNG740 supports these concepts, in this work we measure the correlation between galaxy shapes, as described in eqn.~(\ref{eq.ellpiticity}), and their surrounding environment, as quantified by the tidal shear tensor. A catalog of the tidal shear tensor at each galaxy is one of the many data products which were computed on-the-fly and is provided in the {\small SUBFIND-HBT} catalog of MTNG740. The tidal tensor is defined in three dimensions as:
\begin{equation}
    \Phi,_{ij}(\textbf{x}) = \frac{\partial^2 \Phi(\textbf{x})}{\partial x_i\, \partial x_j},
\end{equation}
where $\Phi(\textbf{x})$, here expressed in units of $c^2$, satisfies the Poisson equation and is obtained by first calculating the three-dimensional overdensity field, $\delta(\textbf{x})$, in the full MTNG740 volume on an $8192^3$ grid using cloud-in-cell interpolation (CIC). We then Fourier transform and Gaussian smooth this field with 20 different smoothing scales, $\lambda _s$, ranging from $148\,{\rm kpc}$ to $7\,{\rm Mpc}$, equally spaced in log. The smoothing is done in $k$-space, and the tidal tensor is then interpolated to the position of every {\small SUBFIND-HBT} object. Hence, the tidal tensor is computed, as in Z22, through:
\begin{equation}
    \Phi,_{ij}(\textbf{x}) = \frac{3\Omega_m}{2\chi_H^2a}\mathcal{F}^{-1}\left[ \frac{k_ik_j}{|\textbf{k}|^2} \mathrm{exp}\left(-\frac{1}{2} |\textbf{k}|^2 \lambda_s^2\right) \mathcal{F}[\delta(\textbf{x})] \right]
    \label{equ:t_tensor} ,
\end{equation}
where $\textbf{k}$ is the comoving wave vector, $\chi_H \equiv c/H_0$ is the Hubble distance, and $a$ is the cosmic scale factor so that $\Phi,_{ij}(\textbf{x})$ is dimensionless. $\mathcal{F}$ and  $\mathcal{F}^{-1}$ are the discrete Fourier and inverse-Fourier transforms, and $\lambda_s$ is the Gaussian smoothing scale. The final tidal shear tensor at each galaxy's position is then obtained by tri-linear interpolation from the tidal field on the grid. The smoothing of the potential helps to avoid substructures introducing distorting small-scale fluctuations in the tidal field, but it is unfortunately not a priori clear which smoothing scale should be chosen to ensure that the galaxy response as a whole is best captured. We will thus consider a range of different smoothing scales in this work.

\subsection{Intrinsic alignment parameters with the tidal field}

\subsubsection{The linear tidal model}

Gravitational interactions between galaxies and their surrounding tidal fields leading to IAs are a cause of contamination in weak lensing signals. For an elliptical galaxy assumed to be in virial equilibrium via the velocity dispersion of its stars, a surrounding anisotropic tidal field can introduce a perturbation causing the galaxy to adjust itself to a modified dynamical equilibrium state. Even if the effect is small (and typically much smaller than the unperturbed shape of the elliptical galaxy, which can be characterized by order unity ellipticity), the resulting new average shape of many galaxies subjected to the same local tidal field will reflect the orientation of the imposed field. A small anisotropy in the local tidal would thus manifest itself as a systematic perturbation in the second moment tensor of the brightness distribution. 

\cite{2018MNRAS.474.1165P} compute the second-order perturbation of the second moment brightness tensor as 
\begin{equation}
    \Delta q_{ij} \equiv \Tilde{q}_{ij}-q_{ij} = - \frac{1}{2\sigma^2}\Phi_{,ab}(\textbf{r}_0)\,S_{abij}
    \label{eq:delta_q},
\end{equation}
where $\Tilde{q}_{ij}$ is obtained using the perturbed mass-to-light relation and $\textbf{r}_0$ is the galaxy location. \cite{2021MNRAS.505.2594G} define $S_{abij}$ as the elasticity tensor, which can be considered as the susceptibility of a galaxy to change its shape under the influence of tidal gravitational fields. They further show that the dominant scaling of ellipticity with the size of a galaxy is dimensionally consistent with the linear tidal model, $q_{ij}=S_{ijab}\Phi_{ab}$.

Thus, following from the above, we define an intrinsic alignment parameter, $D_{\rm IA}$, to be the constant of proportionality between ellipticity (eqn.~\ref{eq.ellpiticity}) and surrounding tidal shear:
\begin{equation}
    \epsilon =  \frac{q_{xx}-q_{yy}}{q_{xx}+q_{yy}} + i\frac{2q_{xy}}{q_{xx}+q_{yy}} = -D_{\rm IA}\left(\Phi_{,xx}-\Phi_{,yy} + 2i\Phi_{,xy}\right),  
\end{equation}
which can be further simplified as 
\begin{equation}
        \epsilon = \epsilon_{+} + i \epsilon_{\times} =  -D_{\rm IA}\left(T_+ + iT_{\times}\right)
        \label{eq.linear_model},
\end{equation}
where $T_+ \equiv \Phi_{,xx}-\Phi_{,yy}$ and $T_{\times} \equiv 2\Phi_{,xy}$. We note that the sign convention for $D_{\rm IA}$ is chosen so that the parameter is strictly positive.

For a more comprehensive discussion of the response of a galaxy to large-scale tidal fields and a detailed derivation of the intrinsic alignment parameter $D_{\rm IA}$ we refer the reader to \citet[][Z22]{2022MNRAS.514.2049Z} and \citet{2021MNRAS.505.2594G}.

\subsubsection{The quadratic tidal torquing model}

For spiral galaxies, a different model is commonly adopted to explain intrinsic alignments, because the orientation of spirals is thought to be imprinted by the generation of angular momentum through tidal torques from the surrounding large-scale matter distribution prior to gravitational collapse. The angular momentum vector of the galaxy in turn determines the inclination angle of the galactic disk relative to the line of sight, and thus its measured ellipticity \citep{2000ApJ...532L...5L,2001ApJ...555..106L}. As neighbouring galaxies have been subjected to correlated initial conditions, this will result in correlated angular momentum directions and thus lead to intrinsic shape correlations \citep{1996MNRAS.282..436C, 2001MNRAS.320L...7C}. 

In spiral galaxies, ellipticity is determined by the angle of inclination under which the stellar disk is viewed. Assuming the symmetry axis of the galactic disk coincides with the direction of angular momentum, $\Hat{\textbf{L}}=\textbf{L}/|\textbf{L}|$, of the stellar component, the ellipticity along the $z$-axis is obtained by:
\begin{equation}
    \epsilon=\frac{\hat{L}^2_y - \hat{L}^2_x}{1+\hat{L}^2_z}+i\frac{2\hat{L}_y\hat{L}x}{1+\hat{L}^2_z}
    \label{eq.spiral_eps}
\end{equation}
\citep{2001ApJ...559..552C}.

As previously noted, the angular momentum in a spiral galaxy is generated by tidal gravitational torquing (for a full review, see \citealt{2009IJMPD..18..173S}) of the large-scale structure acting on a protogalactic cloud \citep{1984ApJ...286...38W,1987ApJ...319..575B, 2013ApJ...769...74S}, which is a first-order process in Lagrangian perturbation theory. Using the most general second order relationship for the variance of the angular momentum directions as a function of the tidal shear tensor,
\begin{equation}
    \langle \hat{L}_i\hat{L}_j|\Tilde{\Phi}\rangle = \frac{1+A}{3}\delta_{ij} - A \Tilde{\Phi}_{,ik}\Tilde{\Phi}_{,kj} ,
\end{equation}
\cite{2001ApJ...555..106L} have found an effective description of the random process $p(\hat{L}_i|\Phi_{,ij})$ that carries the dependence on the orientation of the traceless, unit normalised tidal shear, $\Tilde{\Phi}_{,ij} = \hat{\Phi}_{,ij}/(\hat{\Phi}_{,ij}\hat{\Phi}_{,ij})^{1/2}$ where $\hat{\Phi}_{,ij}\coloneqq\Phi_{,ij} - \frac{1}{3}\Phi_{,kk}\delta_{ij}$ is the traceless part of the tidal shear tensor and has the properties $\Tilde{\Phi}_{,ii}=0$ and $\Tilde{\Phi}_{,ij}\Tilde{\Phi}_{,ji}=1$.

Combining the description of angular momentum as a random process  $p(\hat{L}_i|\Phi_{,ij})$ and the relation between observed ellipticity and angular momentum in eqn.~(\ref{eq.spiral_eps}), we obtain an expression for the complex ellipticity of spiral galaxies as a function of tidal shear, originally derived by \cite{2001ApJ...559..552C} and revised in \cite{2017MNRAS.470.3453S}, in the form of a quadratic model:
\begin{equation}
    \epsilon = \frac{A_{\rm{IA}}}{2}\left(\Tilde{\Phi}_{,xi}\Tilde{\Phi}_{,ix}-\Tilde{\Phi}_{,yi}\Tilde{\Phi}_{,iy} - 2i\Tilde{\Phi}_{,xi}\Tilde{\Phi}_{,iy}\right) \equiv A_{\rm{IA}}\left(Q_+ - iQ_{\times}\right)
    \label{eq.quadratic},
\end{equation}
where $Q_+\equiv 1/2(\Tilde{\Phi}_{,xi}\Tilde{\Phi}_{,ix}-\Tilde{\Phi}_{,yi}\Tilde{\Phi}_{,iy})$ and $Q_{\times}\equiv\Tilde{\Phi}_{,xi}\Tilde{\Phi}_{,iy}$. Here $A_{\rm{IA}}$ absorbs the description of the randomness of the angular momentum field, the misalignment between galactic disk and angular momentum, and the effect of a stellar disk of finite thickness, which further reduces the observed ellipticity with increasing inclination angle.

\begin{figure}
  \includegraphics[width=0.99\columnwidth]{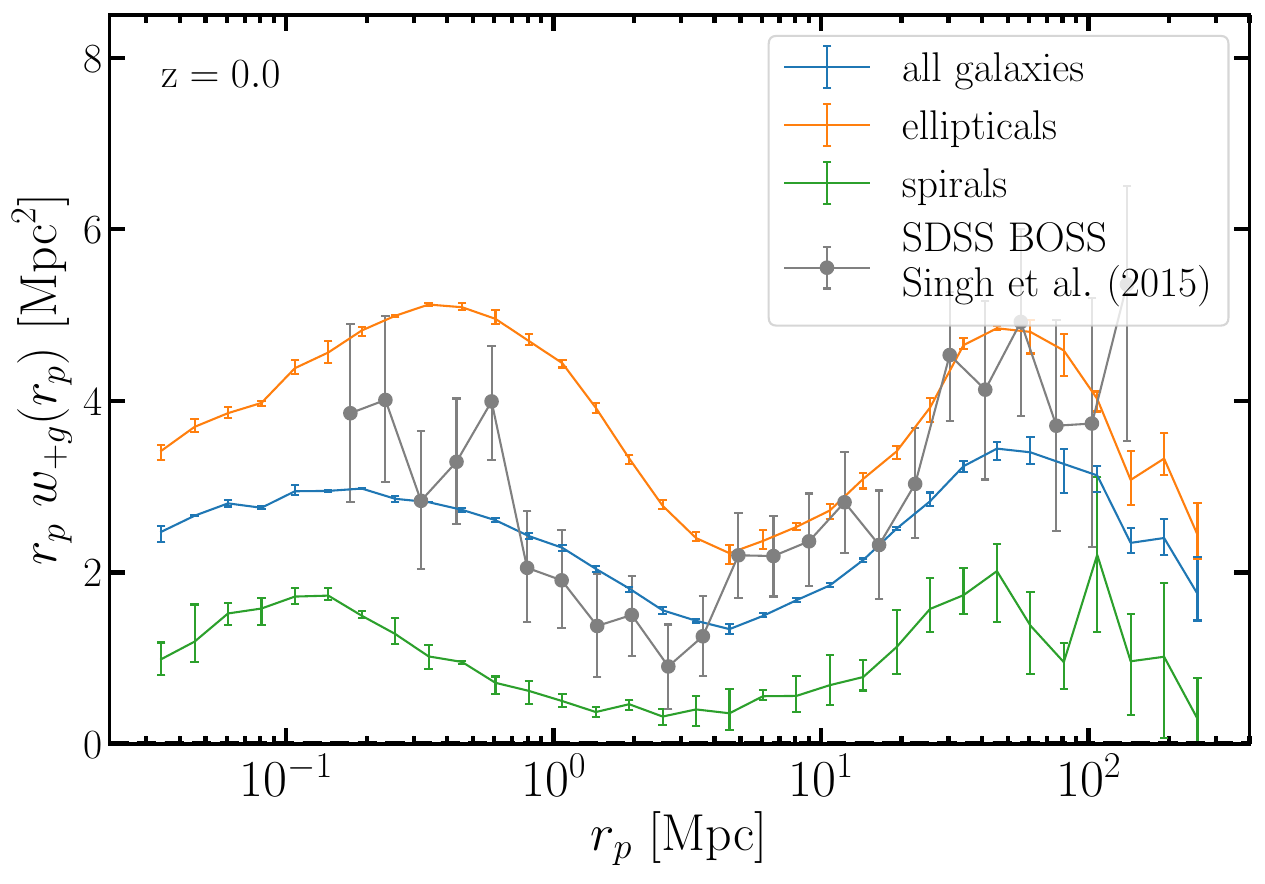}
  \includegraphics[width=0.99\columnwidth]{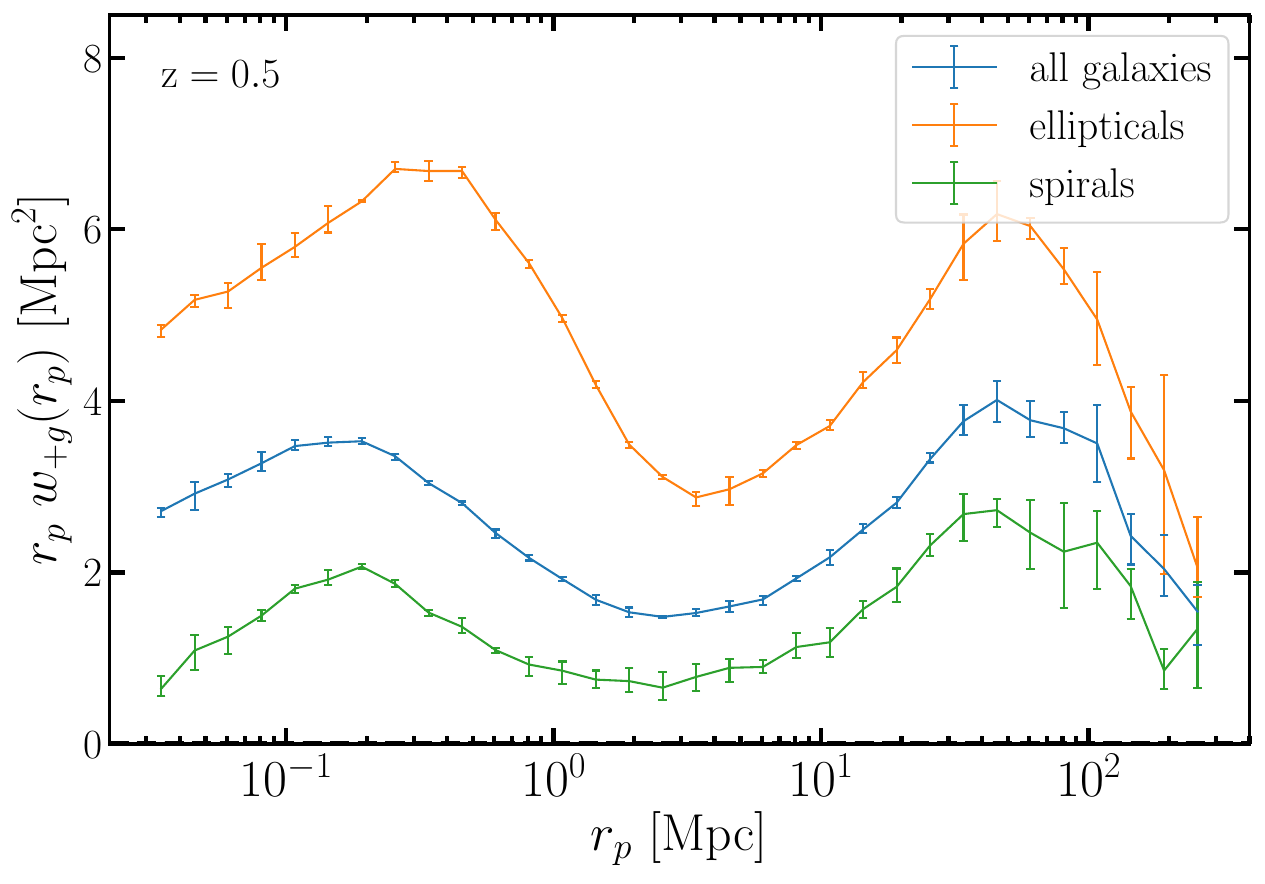}
  \includegraphics[width=0.99\columnwidth]{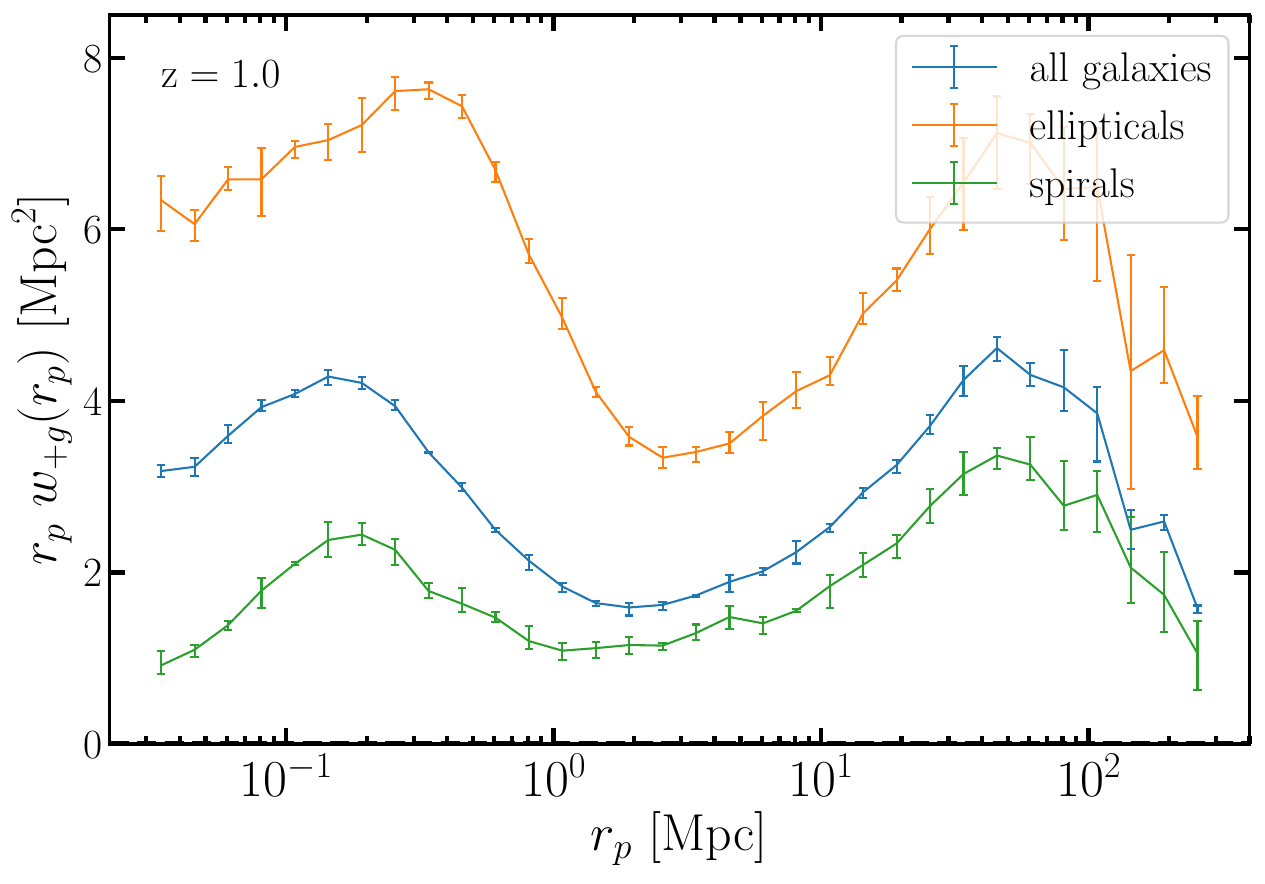}
  \caption{Projected cross-correlations of the intrinsic galaxy shear with galaxy positions, $w^{+g}_{\rm p}(r_{\rm p})$, at three redshifts, $z=0.0$, $z=0.5$, and $z=1.0$, as labelled. In each panel, the correlations are shown separately using the shapes of our subsamples of elliptical and spiral galaxies, as well as for all galaxies, while the density field is always sampled by using all galaxies as unweighted tracers. Our measurements are averages for projections along the $z$-, $y$-, and $x$-directions, with the error bars indicating the maximum and minimum range among them for every bin as a measure of statistical uncertainty. In the $z=0$ panel, we include a measurement of intrinsic alignment obtained by \citet{2015MNRAS.450.2195S} for the low-redshift (LOWZ) sample of the SDSS-III BOSS survey.}
  \label{fig:wp_plot}
\end{figure}

\begin{figure}
  \includegraphics[width=0.99\columnwidth]{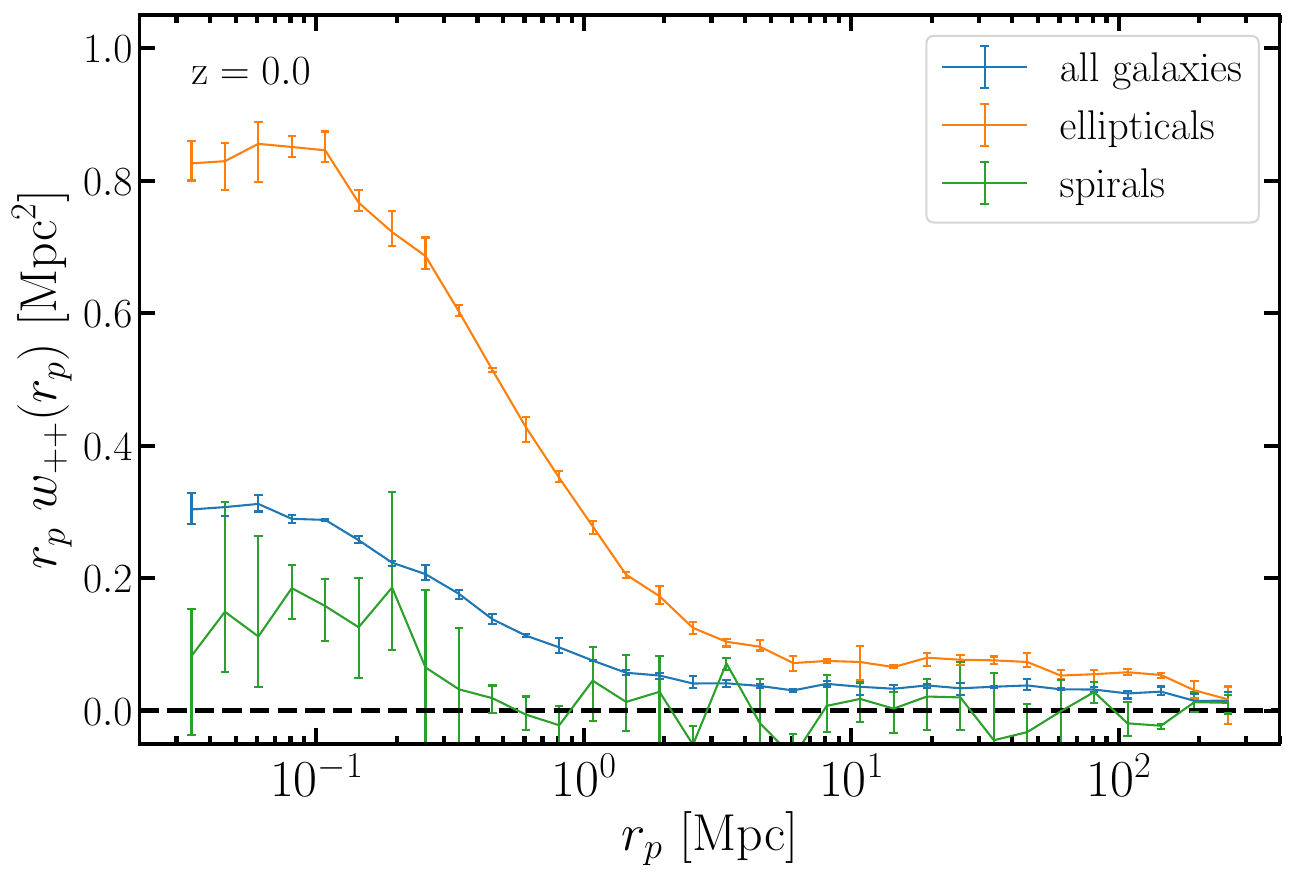}
  \includegraphics[width=0.99\columnwidth]{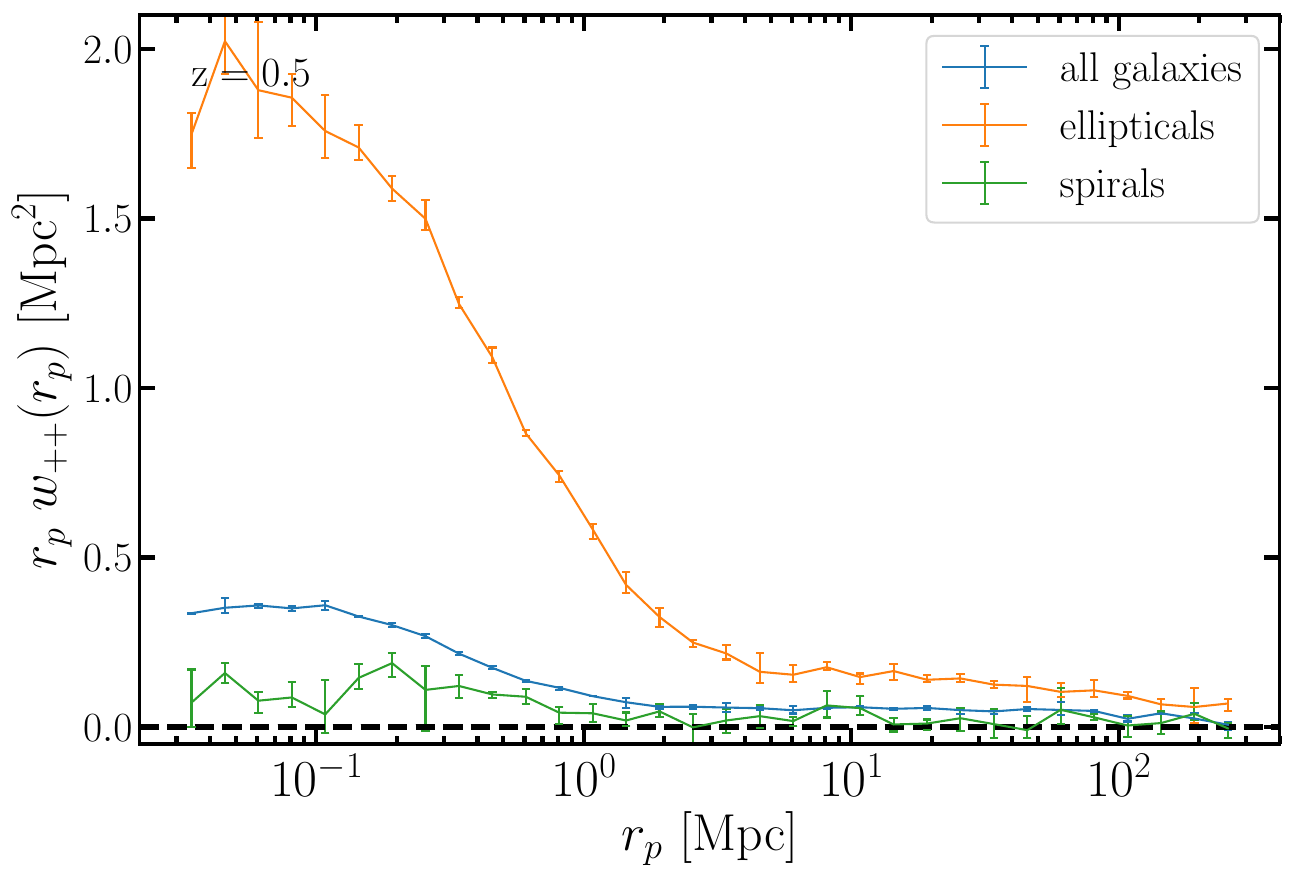}
  \includegraphics[width=0.99\columnwidth]{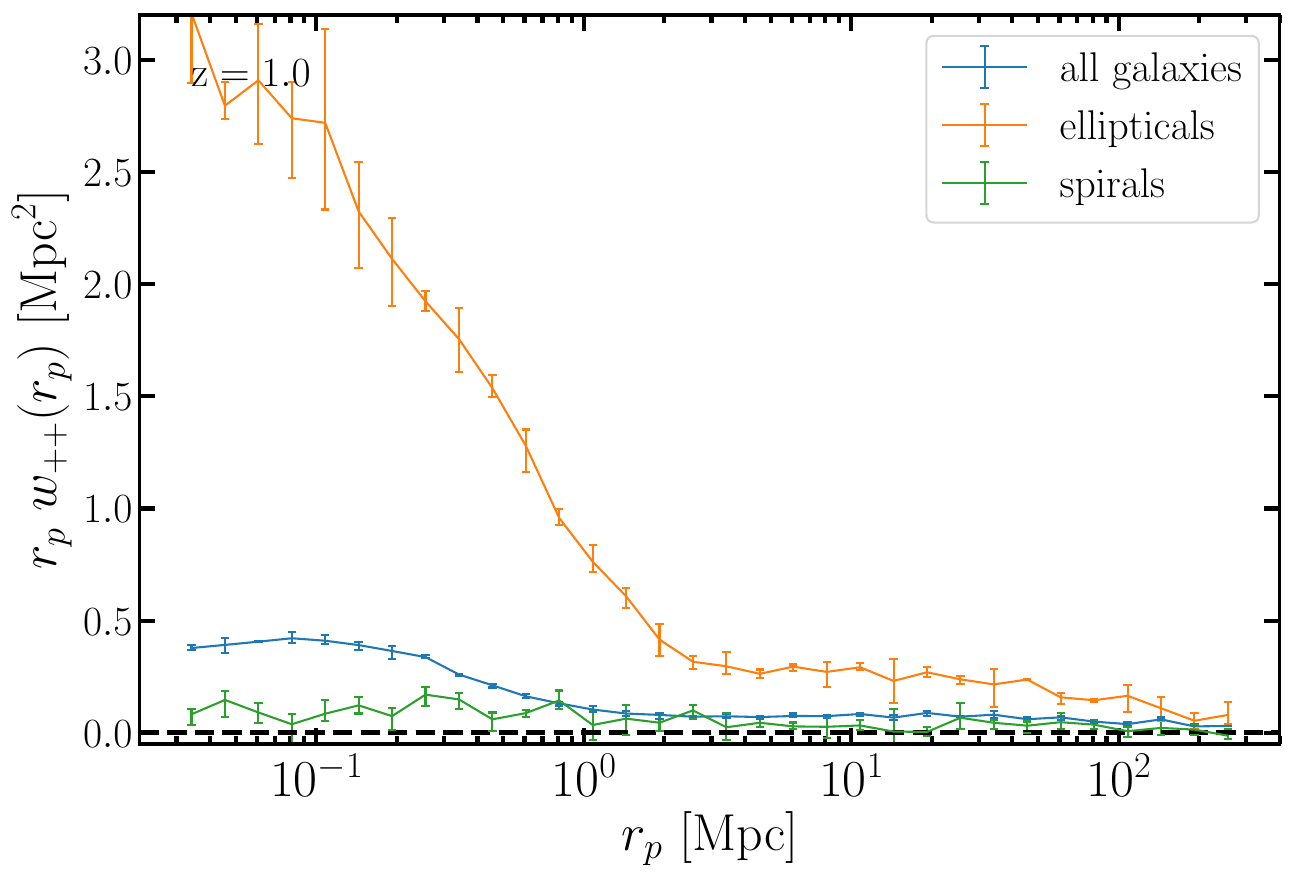}
  \caption{Projected two-point correlation functions of the intrinsic shear-shear correlations of galaxies, $w^{++}_{\rm p}(r_{\rm p})$, at three different redshifts, $z=0.0$, $z=0.5$, and $z=1.0$, as labelled. In each panel, the correlations are shown separately using the shapes of our subsamples of elliptical and spiral galaxies, as well as for our full galaxy sample. We report averages for projections along the $z$-, $y$-, and $x$-directions, with the error bars indicating the maximum and minimum range among them for every bin as a measure of statistical uncertainty. Note that we detect shear-shear correlations for elliptical galaxies with very high significance, all the way up to large distances of $r_{\rm p} \sim 100\,h^{-1}{\rm Mpc}$. In contrast, only weak shear-shear correlations are found for spiral galaxies, for small distances $r \lesssim 1\,h^{-1}{\rm Mpc}$, whereas they are consistent with zero for large separations. }
  \label{fig:wp_shape_shape_plot}
\end{figure}

\begin{figure}
  \includegraphics[width=0.99\columnwidth]{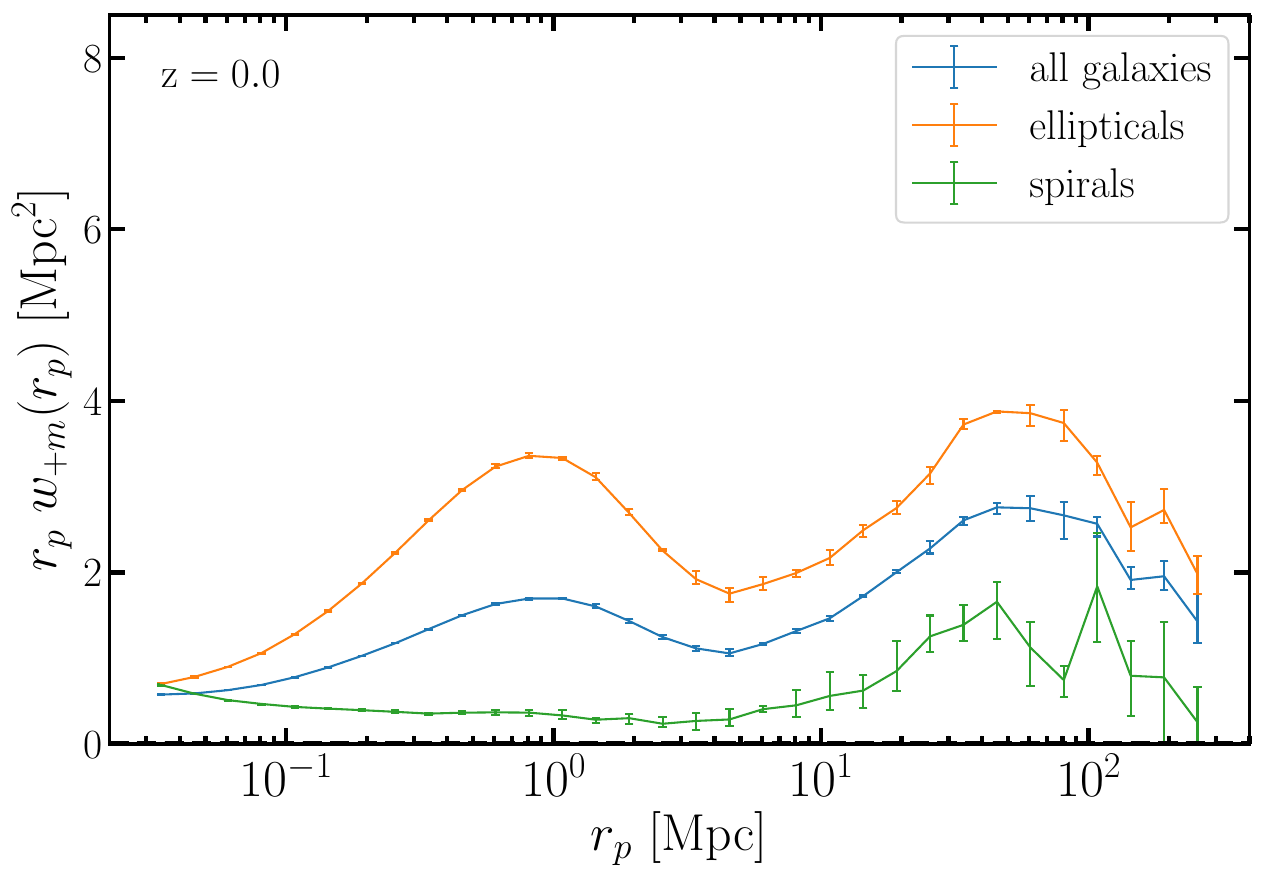}
  \includegraphics[width=0.99\columnwidth]{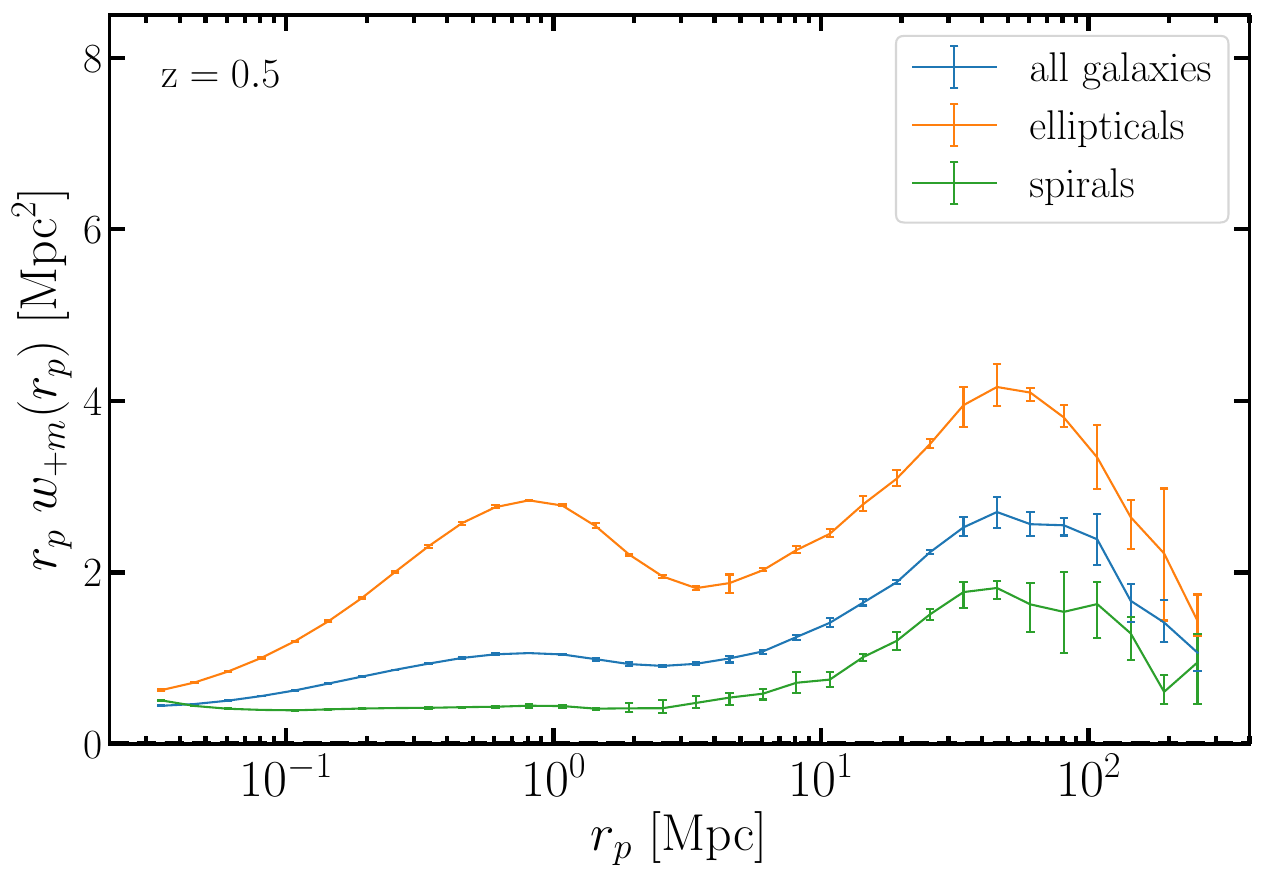}
  \includegraphics[width=0.99\columnwidth]{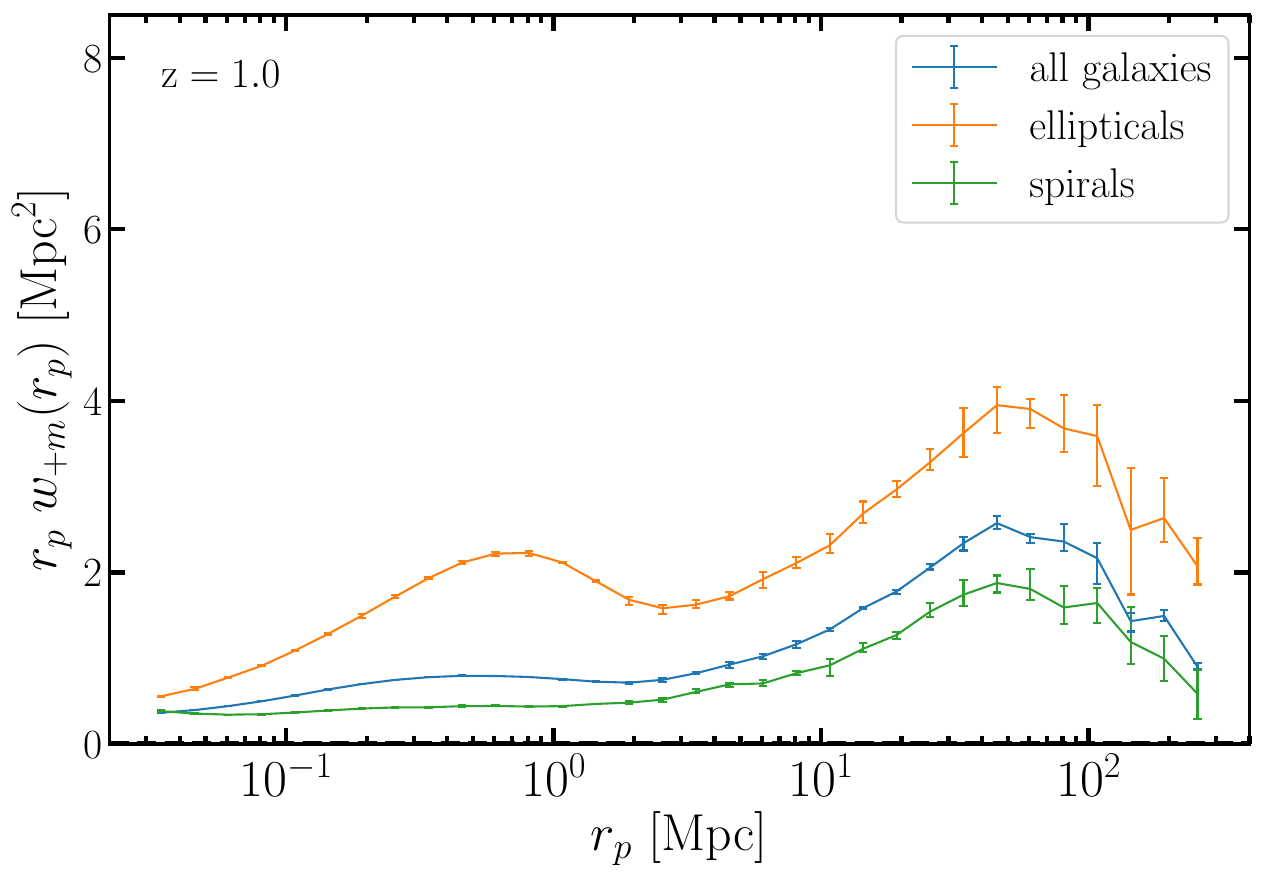}
  \caption{Projected cross-correlations of the intrinsic galaxy shear with the full matter density field of MTNG740, $w^{+m}_{\rm p}(r_{\rm p})$, at three redshifts, $z=0.0$, $z=0.5$, and $z=1.0$, as labelled. In each panel, the correlations are shown separately using the shapes of our subsamples of elliptical and spiral galaxies, as well as for all galaxies, while the density field is directly based on the MTNG740 simulation, and included dark matter, gas, stars, and black holes, appropriately mass weighted. Our measurements are averages for projections along the $z$-, $y$-, and $x$-directions, with the error bars indicating the maximum and minimum range among them for every bin as a measure of statistical uncertainty. }
  \label{fig:wp_shape_matter_plot}
\end{figure}

\begin{figure}   
  \includegraphics[width=0.99\columnwidth]{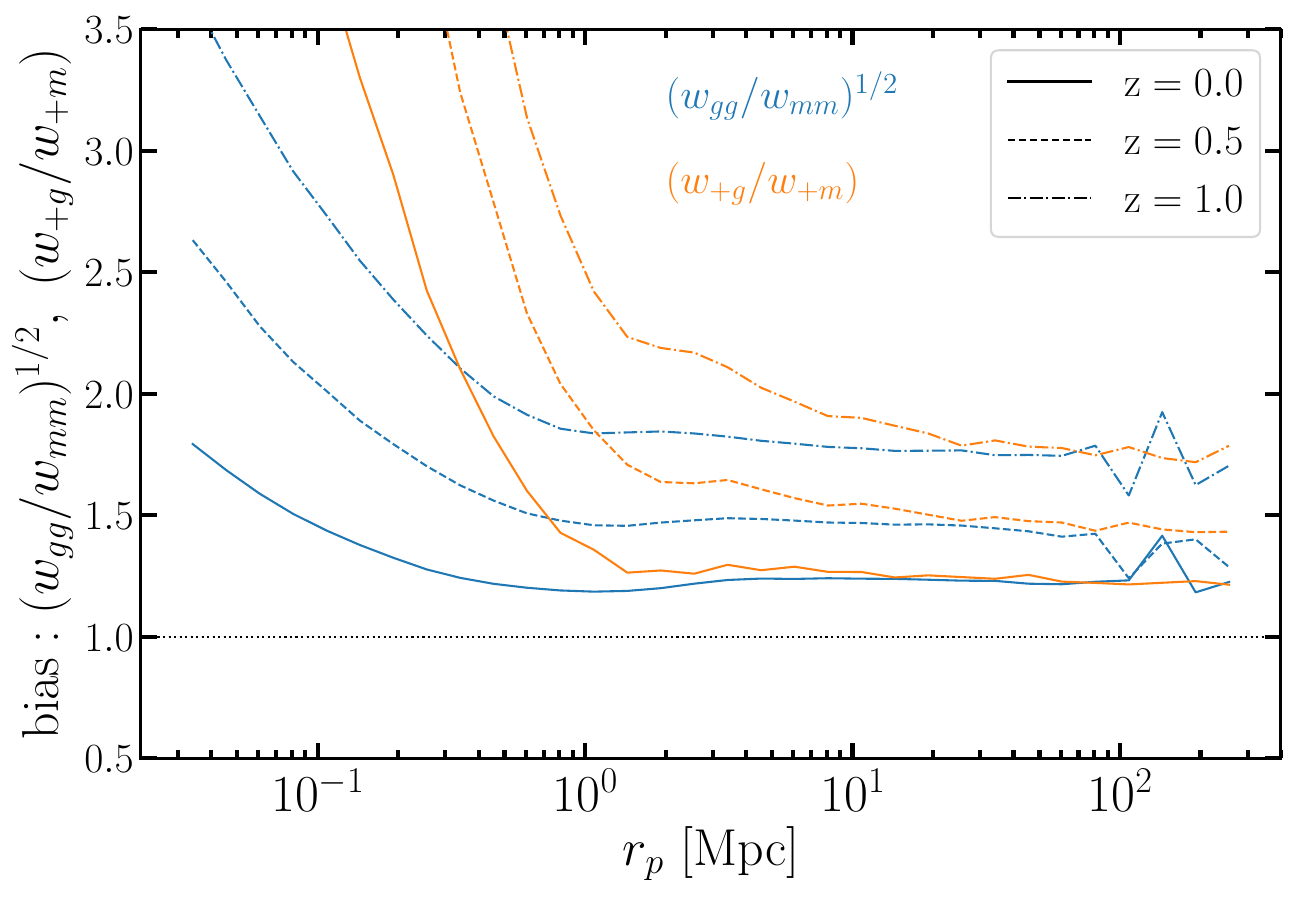}
  \caption{Galaxy bias defined in two different ways, once in the traditional way as $b=(w_{\rm p}^{gg} / w_{\rm p}^{mm})^{1/2}$ based on the auto-correlation functions of galaxies and the full density field, and once as $b'= w_{\rm p}^{+g} / w_{\rm p}^{+m}$ based on the ratio of the correlations of intrinsic shear of galaxies either with themselves or with the matter. We show results for our full galaxy sample, and compare $b$ and $b'$ for three different redshifts. We expect both measures to agree on large scale (but with different values), which is indeed the case for all three redshifts we examined. This shows, in particular, that $w_{\rm p}^{+g}$ and $w_{\rm p}^{+m}$ have the same shapes in the two-halo regime on large scales. In contrast, the fact that $b'$ and $b$ begin to differ for $r_{\rm p}\lesssim 10\,h^{-1}{\rm Mpc}$ demonstrates shape differences between $w_{\rm p}^{+g}$ and $w_{\rm p}^{+m}$ on small scales  and points towards a more complicated relation between shape alignments and density field there, particularly in the one-halo regime. }
  \label{fig:wp_bias_plot}
\end{figure}

\section{Correlations of intrinsic galaxy shear}
\label{results:correlations}

In Fig. \ref{fig:wp_plot}, we present our results for the correlations of intrinsic galaxy shear with galaxy positions, $w^{+g}_{\rm p}(r_{\rm p})$, at three redshifts, $z=0.0$, $z=0.5$, and $z=1.0$, as labelled. In each panel, the correlations are shown separately for MTNG740 using the shapes of our subsamples of elliptical and spiral galaxies, as well as for all galaxies, while the density field is always sampled by using all galaxies as unweighted tracers. Note that these results take full advantage of the  statistical power of our direct hydrodynamical simulation of galaxy formation, and report the intrinsic alignment signal over a much larger radial range and with much smaller statistical uncertainty than ever reported in the literature thus far. Note that we plot the quantity $r_{\rm p} w^{+g}_{\rm p}$ instead of $w^{+g}_{\rm p}$ in order to compress the vertical dynamic range, allowing us to use a linear axis for $r_{\rm p} w^{+g}_{\rm p}$.

In the $z=0.0$ panel of Fig.~\ref{fig:wp_plot}, we include a measurement of intrinsic alignment obtained by \citet{2015MNRAS.450.2195S} for the low-redshift (LOWZ) sample of the SDSS-III BOSS survey. Interestingly, these observational findings match the amplitude and shape of our signal quite well, and the sizes of  residual differences is so small that they could be easily explained, for example, by differences in galaxy sample definition. A similar level of consistency is obtained for recent measurements of IA for luminous red galaxies in the  Kilo-Degree Survey \citep[KiDS-1000,][]{2021A&A...654A..76F}. We consider this good agreement to be extremely encouraging for the use of MTNG740 to provide accurate ``ab-initio'' predictions for the effects of intrinsic alignment on weak lensing statistics.  

We note that our results differ on large scales from those reported by  \cite{2015MNRAS.454.2736C} for the Horizon-AGN simulation, who found a weak correlation between galaxy shapes and positions in the two-halo term, with an amplitude considerably smaller than reported here. In contrast, our results are in somewhat better agreement with those of \citet{2021MNRAS.508..637S}, who measured and compared intrinsic alignments for the three hydrodynamical simulations Illustris, MassiveBlack-II, and TNG300. They found significant IA signals for $w^{+g}_{\rm p}$ for all three simulations, albeit with substantial scatter between the simulation models, and only for a smaller radial range over which IA signals could be detected with statistical significance. Typically, the amplitudes of their reported IA signals are noisier and a bit lower than the ones we find here,  in part reflecting the much smaller box sizes of the simulations they analyzed. 

In Fig. \ref{fig:wp_shape_shape_plot}, we show our results for the direct correlations of the intrinsic shear of galaxies, $w_{\rm p}^{++}$. Interestingly, we find a  clear and statistically significant alignment signal for elliptical galaxies even out to distances as large as $r_{\rm p}\simeq 100\,h^{-1}{\rm Mpc}$, at all redshifts. On small scales corresponding to the one-halo regime, $r_{\rm p}\lesssim 1\,h^{-1}{\rm Mpc}$, the signal is strongly rising for ellipticals, until $r_{\rm p}\,w_{\rm p}^{++}$ appears  to reach a plateau at distances below $\sim100\,h^{-1}{\rm kpc}$. Interestingly, there is also a significant redshift evolution in the amplitude of the shear-shear correlations of ellipticals, with significantly stronger correlations at  $z=1.0$ than at $z=0.0$. However, as we will show below, this is driven by the evolution of the bias with redshift rather than a change in the effective strength of galaxy alignments.

In contrast, the shear-shear correlations for spirals are much weaker and consistent with zero for $r_{\rm p}\gtrsim 1\,h^{-1}{\rm Mpc}$, although in the one-halo regime we do find evidence for a weakly positive $w_{\rm p}^{++}$ for spirals. This signal shows little, if any, redshift evolution.

\begin{figure}   
  \includegraphics[width=0.99\columnwidth]{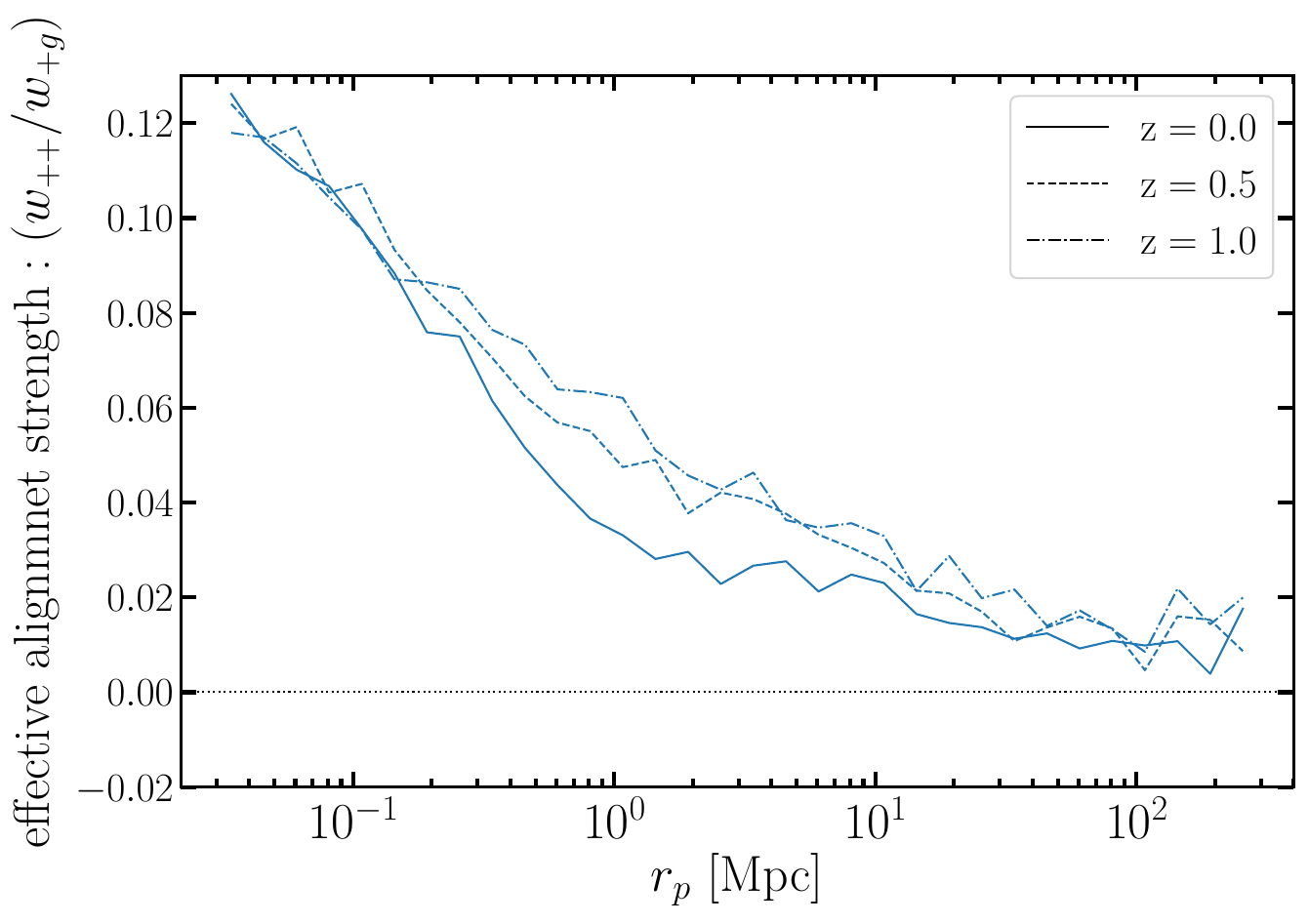}
  \caption{Effective alignment strength for our full galaxy sample, defined as $\alpha(r_{\rm p}) = w_{\rm p}^{++} /  w_{\rm p}^{+g}$; i.e.~the ratio of intrinsic shear-shear correlations to the shear-galaxy correlations. We here show results for our full galaxy sample, and compare the three redshifts $z=0.0$, $0.5$, and $1.0$. This quantity evolves smoothly between the one-halo and two-halo regimes, and is nearly invariant with redshift over the range we examined, modulo a small change in the transition regime $r_{\rm p}\simeq 1\,h^{-1}{\rm Mpc}$.}
  \label{fig:wp_effectivestrength_plot}
\end{figure}

Finally, in Fig. \ref{fig:wp_shape_matter_plot}, we turn to intrinsic shape correlations with the full matter field, $w_{\rm p}^{+m}$. These measurements are based on the density distribution of the hydrodynamical simulation, taking into account both the dark matter and baryonic components.  Again, we find high signal-to-noise radial intrinsic alignments both for ellipticals and spirals over the full radial range that we measured. Note that  the amplitude of $w_{\rm p}^{+m}$  declines towards higher redshift, which is opposite to the redshift evolution we found for $w_{\rm p}^{+g}$ which actually showed an increase towards higher redshift. In fact, the evolution of $w_{\rm p}^{+m}$ in the two-halo regime can be understood well simply with the linear growth factor; i.e.~the shape of $w_{\rm p}^{+m}$ is very similar at $z=0.0$, 0.5, and 1.0 when scaled with linear theory. This underlines that the very different redshift evolution of $w_{\rm p}^{+g}$ and $w_{\rm p}^{+m}$ originates in a rapidly evolving galaxy bias, and that assessing the strength of intrinsic alignment signals with galaxies in weak lensing analysis pipelines thus requires a very precise understanding of galaxy bias.

To make this point more explicit, we consider in Fig. \ref{fig:wp_bias_plot} the galaxy bias for our full galaxy sampled, defined as $b = (w_{\rm p}^{gg} / w_{\rm p}^{mm})^{1/2}$ in terms of the projected two-point correlations of galaxies ($w_{\rm p}^{gg}$) and the projected autocorrelation function of all matter ($w_{\rm p}^{mm}$), both of which we have also measured in addition to the quantities discussed earlier. Using our intrinsic shear correlations we can also define a different bias measure,  $b' = w_{\rm p}^{+g} / w_{\rm p}^{+m}$. We expect that $b'$ should be similar to $b$ at least on large scales, because the `+' sample in the two correlation functions appearing in  $b'$ should enter with an effective alignment strength that then cancels in the ratio, so that only one factor of the bias remains. The results of Fig.~\ref{fig:wp_bias_plot} show that this expectation is borne out in the two-halo regime at large scales. We there obtain $b'=b$, but with different values for each of the examined redshifts, as expected. This shows that $w_{\rm p}^{+g}$ and $w_{\rm p}^{+m}$ have the same shapes at large separations. Conversely, the disagreement of $b'$ and $b$ in the transition region and in the one-halo regime shows that the shapes of these two correlation functions are noticeably different.

We may also consider another interesting ratio of the correlation functions in the form of $\alpha = w_{\rm p}^{++} /  w_{\rm p}^{+g}$. Because the effective alignment strength of the galaxies enters quadratically in the numerator but only linearly in the denominator, whereas  the bias parameter enters quadratically in both, the quantity $\alpha$ should be a measure of effective alignment strength of the galaxies as a function of distance, and be largely independent of the usual bias parameter. In Fig. \ref{fig:wp_effectivestrength_plot}, we show this quantity for our full galaxy sample, comparing all three redshifts we have considered. Interestingly, $\alpha(r_{\rm p})$ declines smoothly from small to large distances, without any obvious break between the one-halo and two-halo regimes. Furthermore, this quantity shows hardly any evolution with redshift, aside from a small reduction with time at distances around $r_{\rm p}\simeq 1\,h^{-1}{\rm Mpc}$.

\section{Alignments of galaxies with the local tidal field}
\label{IA_gals}
In this section we present our results on the intrinsic alignment of galaxies at $z=0.0,\ z=0.5$ and $z=1.0$, as measured by the parameter \DIA~defined in eqn.~(\ref{eq.linear_model}). We identified our galaxy morphology based on colour and rotational support as described in Section~\ref{Methods:Galaxies}.

\subsection{Elliptical galaxies}
\label{results:IA_ells}

\begin{figure*}
    \centering
    \includegraphics[width=0.95\textwidth]{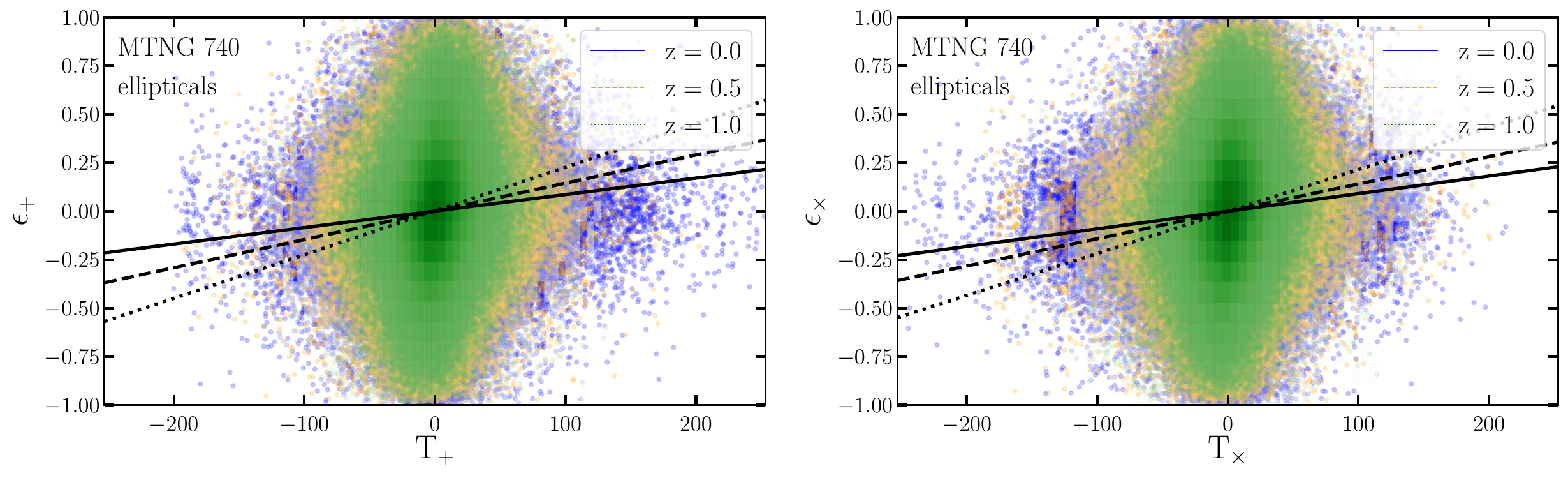}
    \includegraphics[width=0.95\textwidth]{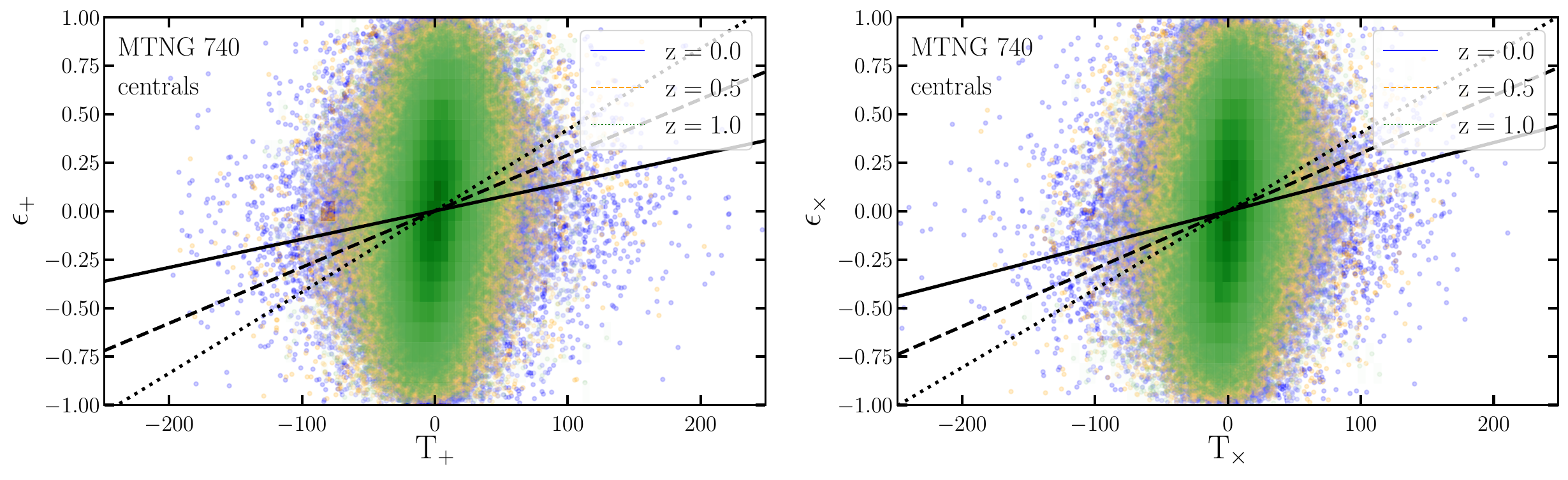}
    \caption{Correlations of the ellipticity of elliptical galaxies and the tidal field. The left panels show correlations for the tangential components, the right plot for the cross components. The bottom two panels are a subset of the top two panels and contain a selection of central galaxies only. Results are shown for redshifts $z=0.0$ (blue region, solid black line), $z=0.5$ (orange region, dashed black line) and $z=1.0$ (green region, dotted black line). Each panel is overlaid with the 2D histogram at $z=1.0$ to provide the reader with an idea of the galaxy distribution: we find a high concentration of galaxies in the centers of the distributions. The tidal tensor is calculated at a smoothing scale of 7~Mpc.  Fitting was done on the mean values of ellipticity within bins such that there are 1,000 galaxies in each bin. We define the intrinsic alignment strength \DIA~as the value of the slope. We see a stronger \DIA~signal at $z=1.0$.}
    \label{fig:IA_ellipticals}
\end{figure*}

Fig. \ref{fig:IA_ellipticals} shows the real and imaginary parts of the observed galaxy ellipticity in the $V$-band vs.~its corresponding tidal field value at a smoothing scale of $\lambda_s = 7\,{\rm  Mpc}$. Each data point represents an individual galaxy, and we overlay our three redshift galaxy selections. Elliptical galaxies at $z=0.0$ appear in blue, those at $z=0.5$ appear in orange, and the green data points pertain to elliptical galaxies at $z=1.0$.
 
We bin the selected galaxies so that there are 1,000 galaxies in each bin. The intrinsic alignment parameter, \DIA, is then measured by performing linear regression of the mean ellipticity in each bin as a function of the central value of each tidal tensor bin, $\epsilon_+(T_+)$ and $\epsilon_{\times}(T_{\times})$, as per eqn.~(\ref{eq.linear_model}). It is worth noting that the \DIA~ measurement is sensitive to the method in which the galaxies are binned. The top two panels of Fig.~\ref{fig:IA_ellipticals} further illustrate the fitting lines from the linear regression. We see that there is an increased alignment signal at higher redshift. We also emphasize that \DIA$_+$ and \DIA$_\times$ are expected to give consistent results, which can serve as a useful consistency check.

 Z22 performed 1,000 random galaxy and tidal field rotations for each galaxy in their study in order to mitigate the small volume of TNG100. They found that averaging all intrinsic alignment parameter values from their randomisation procedure was consistent with the value obtained by simply averaging  their \DIA$_+$ and \DIA$_{\times}$ components. At $z=0.0$, we report an intrinsic alignment parameter of the tangential component of observed elliptical galaxies in the V-band from MTNG740 with value \DIA$_+ = (8.51\pm0.29)\times10^{-4}$ and that of the transverse component as \DIA$_{\times} = (9.08\pm0.28)\times10^{-4}$. These intrinsic alignment signals are in close agreement (with only a 6.5\% difference between the two) and are illustrated as the solid black lines in the top two panes of Fig.~\ref{fig:IA_ellipticals}. Averaged together they give an intrinsic alignment parameter value of
\begin{equation}
\label{equ:DIA_z0.0}
    D_{\rm{IA}(z=0.0)}=(8.79\pm0.20)\times10^{-4},
\end{equation}
which is $\sim44\,\sigma$ different from zero.

The bottom two panels in Fig.~\ref{fig:IA_ellipticals} are the same as the top two panels but show the projected ellipticity as a function of the tidal field value for elliptical galaxies that are also identified as central galaxies in their host halos. This is a significantly smaller galaxy sample than all ellipticals, containing $\sim$300,000 centrals. We see stronger intrinsic alignments of central galaxies for both tangential and transverse components as compared to all ellipticals. The average parameter value for centrals is $D_{\rm{IA}(z=0.0)}=(1.61\pm0.05)\times10^{-3}$, which is about 2 times higher than that of all ellipticals. It is reasonable to assume that galaxies that are not centrals in their host halos are more susceptible to perturbing events, such as recent mergers, which can disrupt any intrinsic alignment they may have with their surrounding tidal shear field. It is therefore unsurprising that central galaxies show stronger intrinsic alignment strength than does a sample of all elliptical galaxies.

At $z=0.5$, we report an intrinsic alignment parameter of the tangential component of elliptical galaxies with a value \DIA$_+ = (1.45\pm0.05)\times10^{-3}$ and that of the transverse component as \DIA$_{\times} = (1.41\pm0.05)\times10^{-3}$. Again, the values are in close agreement ($\sim$3.1\% difference) and are illustrated in Fig. \ref{fig:IA_ellipticals} as black dashed lines. The average of these two components gives an intrinsic alignment value of
\begin{equation}
\label{equ:DIA_z0.5}
    D_{\rm{IA}(z=0.5)}=(1.43\pm0.04)\times10^{-3},
\end{equation}
which is about $36\,\sigma$ different from zero and 1.6 times stronger than $D_{\rm{IA}(z=0.0)}$ for ellipticals.

At $z=1.0$, we obtain an intrinsic alignment parameter for the tangential component of elliptical galaxies with a value of \DIA$_+ = (2.25\pm0.09)\times10^{-3}$ and that of the transverse component as \DIA$_{\times} = (2.16\pm0.09)\times10^{-3}$. Again, these two values are in close agreement ($\sim$4.1\% difference), and they are illustrated in Fig.~\ref{fig:IA_ellipticals} as black dotted lines. The average of the two components gives an intrinsic alignment value of
\begin{equation}
\label{equ:DIA_z1.0}
    D_{\rm{IA}(z=1.0)}=(2.21\pm0.06)\times10^{-3},
\end{equation}
which is still significantly different from zero at about 37$\sigma$, and is 2.5 times stronger than $D_{\rm{IA}(z=0.0)}$ for ellipticals. 

Our results are consistent with previous studies that showed larger intrinsic alignment of elliptical galaxies at higher redshifts \citep{2015MNRAS.448.3522T, 2020ApJ...904..135Y, 2021MNRAS.508..637S, 2022MNRAS.514.2049Z}. We emphasize, however, that the linear alignment model itself does not have a redshift dependence. The larger intrinsic alignment signal at higher redshifts is mostly a result of the change in galaxy sample as we move to earlier times. We explore the corresponding scaling of \DIA with galaxy mass and the tidal smoothing scale  in Section~\ref{sec:environment}.  

\subsection{Spiral galaxies}
\label{results:IA_spirals}

\begin{figure*}
    \centering
    \includegraphics[width=0.95\textwidth]{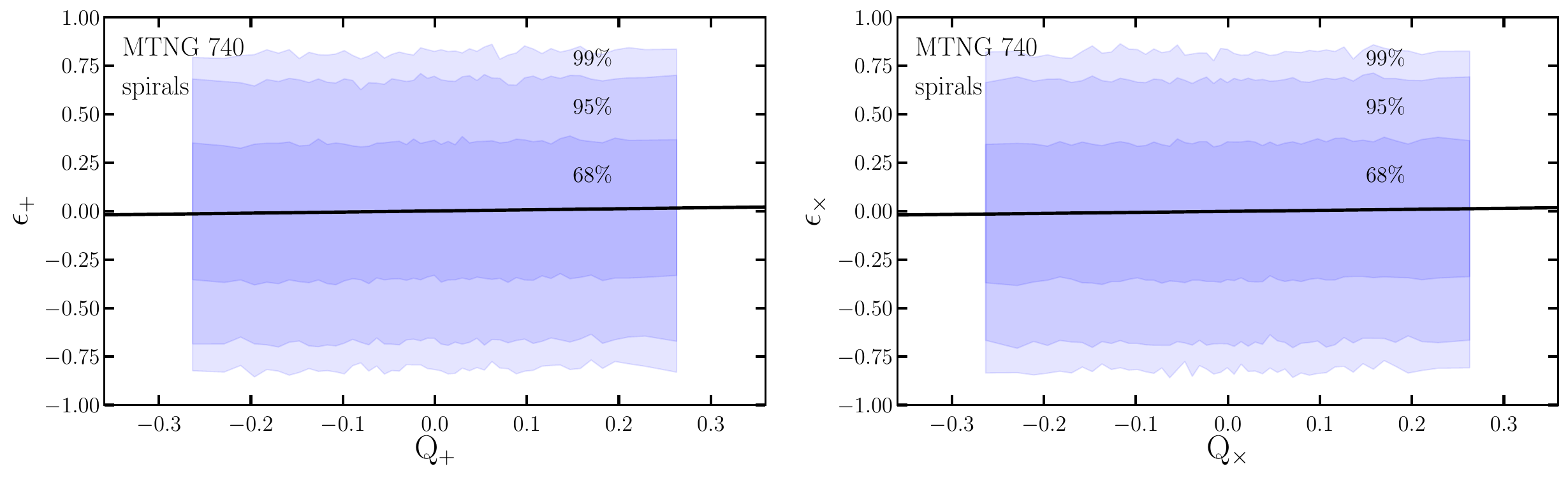}
        \caption{In contrast to what is shown in Fig.~\ref{fig:IA_ellipticals}, intrinsic alignment of spiral galaxies under a quadratic model, shown here, is significantly weaker. We depict the inner percentile ranges of the distribution of spiral galaxies under a quadratic model (blue shaded regions) and slope representative of $A_{\rm IA}$ (black solid line). Results at all three redshifts are consistent so for clarity we show only $z=0.0$. 
        }
    \label{fig:IA_spirals}
\end{figure*}

We select our spiral galaxies using the method described in Section~\ref{Methods:Galaxies}. As illustrated in the rightmost panel of Fig.~\ref{fig:galaxy_selection}, this approach, along with the initial mass cuts in galaxy selection, yielded a sample of spiral galaxies that is significantly smaller than that of our elliptical galaxies.

Fig.~\ref{fig:IA_spirals} shows the real and imaginary components of the observed spiral galaxy ellipticity  as a function of the corresponding tidal field component value again at a smoothing scale of $\lambda_s = 7\,{\rm  Mpc}$. As in Fig.~\ref{fig:IA_ellipticals}, we overlay results for three redshifts ($z=0.0$, $z=0.5$, and $z=1.0$) in each panel. The two panels show our results from fitting the quadratic model. 

Evidently, there is rather substantial scatter between the measured ellipticities and the parameters of the tidal torquing model. Still, we have good enough statistics to discern a correlation nevertheless. In fact, we find scores of $A_{\rm{IA}}$ at all redshifts for intrinsic alignments of spiral galaxies under a quadratic model. Both real and imaginary parts of the ellipticity vs.~shear tensor $\epsilon_{+,\times}(Q_{+,\times})$ are in agreement with the real and imaginary component scores averaged together for values of
\begin{subequations}
\begin{align}
    &A_{\rm{IA}(z=0.0)}=0.056\pm0.007,\\
    &A_{\rm{IA}(z=0.5)}=0.047\pm0.004,\\
    &A_{\rm{IA}(z=1.0)}=0.070\pm0.005.
\end{align}
\end{subequations}
However, these values show only a moderately significant difference from zero compared to elliptical galaxies, formally corresponding to  $8\,\sigma$, $12\,\sigma$ and $14\,\sigma$, at redshifts $z=0.0$, $0.5$ and $1.0$, respectively. 

Unlike Z22 who found a significant quadratic signal only for the most massive spirals at $z = 1$, we are thus able to identify this correlation at all redshifts, albeit in a considerably less clear fashion and weaker strength than for the linear model applied to elliptical galaxies. This corroborates doubts about the practical utility of the tidal torquing model and questions whether the orientation of spirals today is still in the direction of the angular momentum imprinted on the halo prior to its collapse. For example, cosmological simulations of Milky Way-sized disk galaxies show that they turn on average by 40 degrees between $z=1$ and $z=0$, and thus do not have a particularly stable orientation in time \citep{2015MNRAS.452.2367Y}.

\subsection{Dependence on galaxy and environment}
\label{sec:environment}

\begin{figure}
    \centering
    \includegraphics[width=0.95\columnwidth]{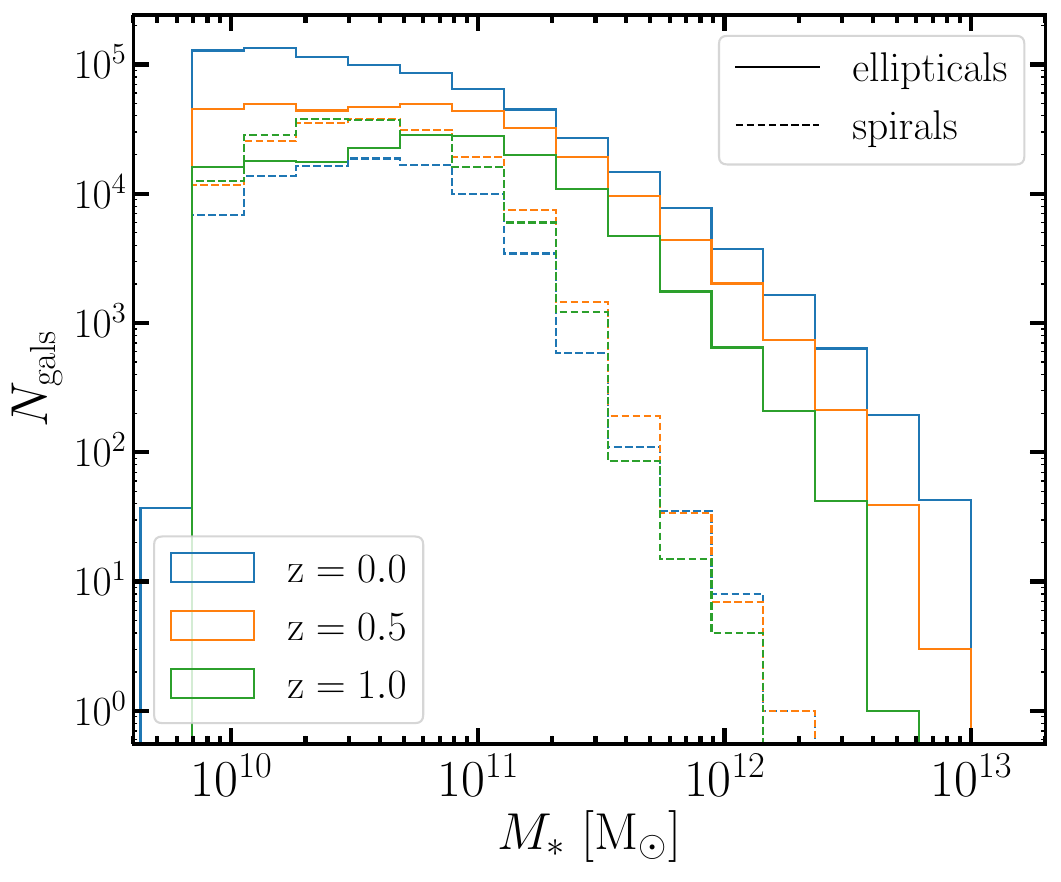}
    \caption{We show a histogram of our morphological split at  $z=0.0$, $z=0.5$, and $z=1.0$ and binned by stellar mass.}
    \label{fig:Ngals}
\end{figure}

\begin{figure*}
    \includegraphics[width=0.99\columnwidth]{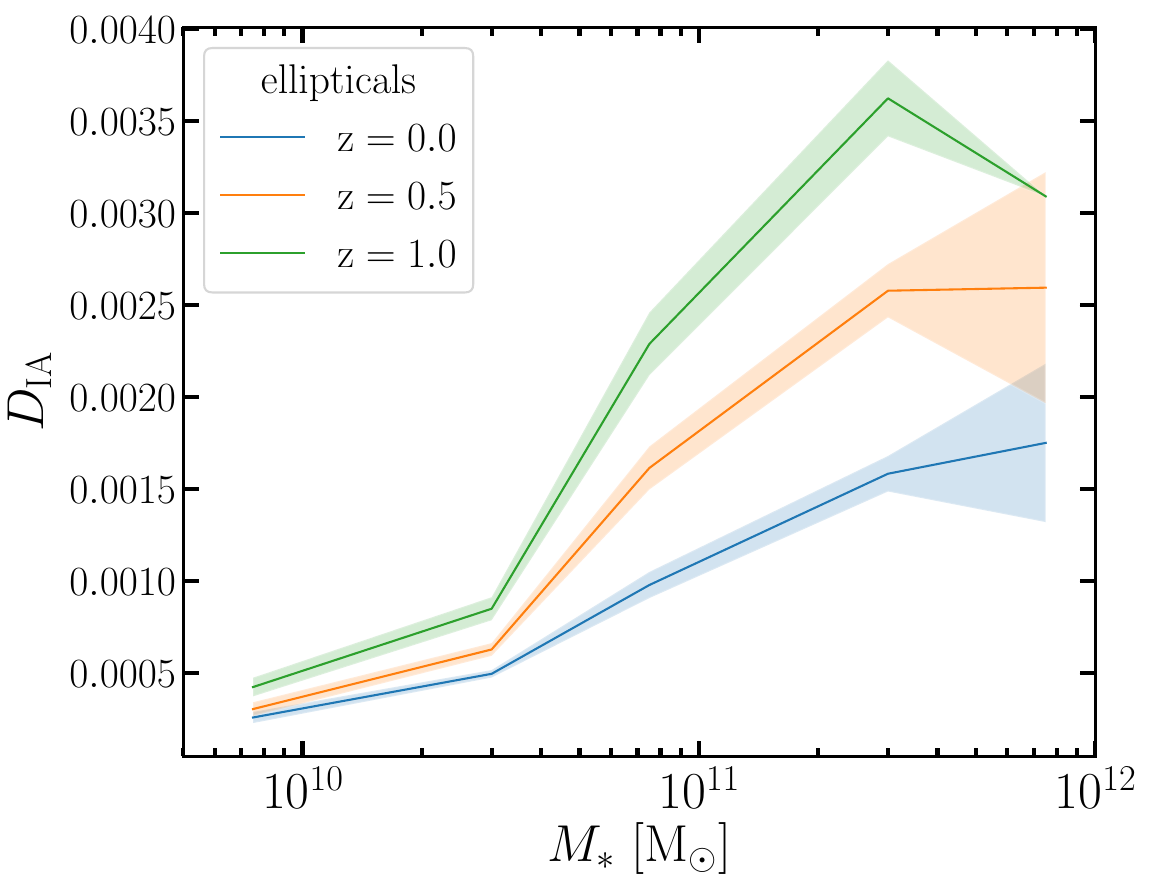}
    \includegraphics[width=0.95\columnwidth]{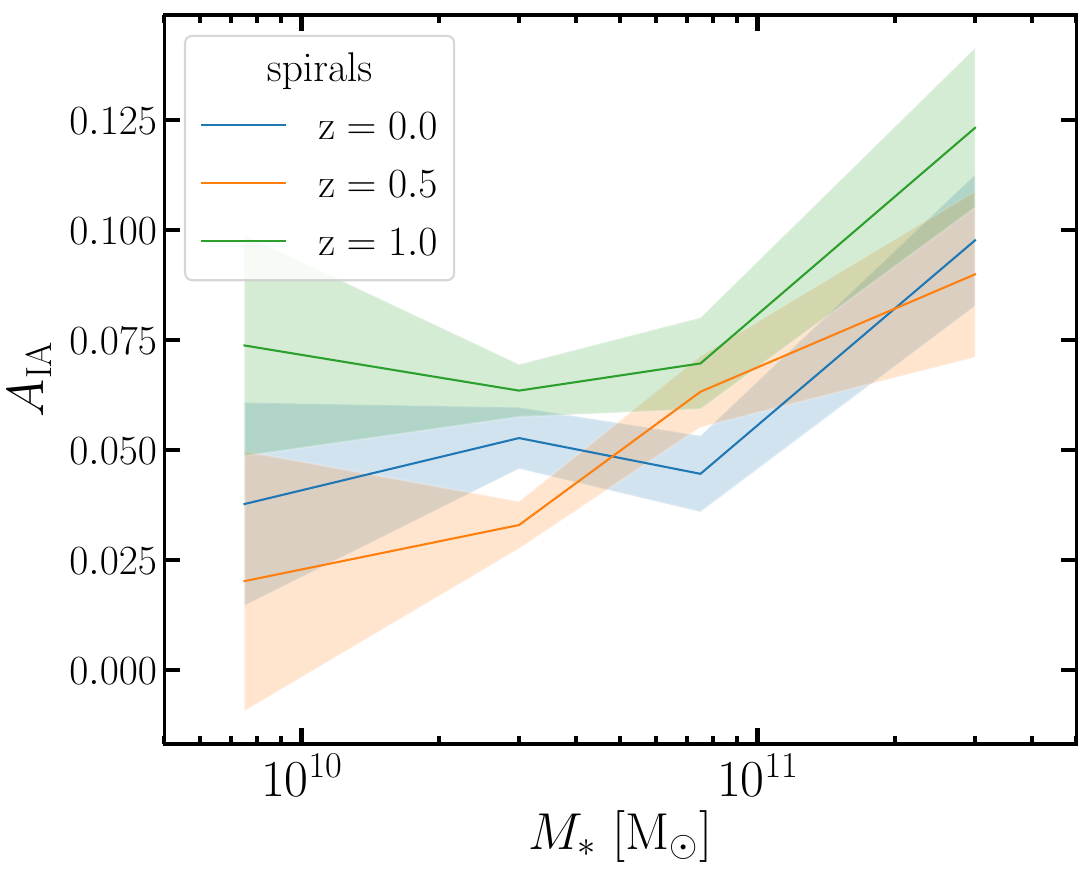}
    \caption{Intrinsic alignment strengths as a function of stellar mass. Results are shown for elliptical galaxies (left) and spiral galaxies (right) at three different redshifts: $z=0.0$, $z=0.5$, and $z=1.0$.  The tangential and cross components of the IA strength were found to be highly consistent with each other, these figures therefore show the average IA strengths. There is a clear trend of greater IA magnitude with increased stellar mass.
    }
    \label{fig:IA_starmass}
\end{figure*}

We now examine how the intrinsic alignment strengths, \DIA and $A_{\rm{IA}}$, in elliptical  and spiral galaxies, respectively, depend on galaxy and environmental properties. We first look at the intrinsic alignment strength dependence on stellar mass. Fig.~\ref{fig:Ngals} is a histogram of the number of galaxies as function of stellar mass for elliptical and spiral populations at our three redshifts. We note the overall larger population of elliptical galaxies compared to spiral galaxies. This is expected at higher mass bins as the total galaxy population becomes dominated by ellipticals at higher stellar masses. \cite{2022arXiv221010060P} show that at masses $M_*<10^{10.5}$\Msol, blue galaxies dominate the galaxy population in MTNG740. For this study, however, our galaxy selection relies on being able to resolve galaxies with a minimum number of star particles, in which case MTNG740 does not have the resolution needed for low-mass galaxies, which is where spirals would dominate. We selected a mass range which would result in $N_{\mathrm{gals}}>1000$ galaxies per bin.
To this end, we created 5 stellar mass bins:
\begin{itemize}
    \item $M_* = (5\times10^{9}-10^{10})\,$\Msol
    \item $M_* = (10^{10}-5\times10^{10})\,$\Msol
    \item $M_* = (5\times10^{10}-10^{11})\,$\Msol
    \item $M_* = (10^{11}-5\times10^{11})\,$\Msol
    \item $M_* = (5\times10^{11}-10^{12})\,$\Msol
\end{itemize}
in which we calculated \DIA~in the $V$-band for elliptical galaxies, but only used the first four mass bins to calculate $A_{\mathrm{IA}}$ for spiral galaxies as there are significantly less than 1000 spirals in the highest mass bin. The results are shown in Fig.~\ref{fig:IA_starmass}. There is a general trend of increasing intrinsic alignment strength with increasing stellar mass, in agreement with Z22 and other studies.

Upon initial examination of Fig.~\ref{fig:IA_starmass}, if we were to naively match the values of \DIA~at a given redshift as per equations (\ref{equ:DIA_z0.0}),  (\ref{equ:DIA_z0.5}) and (\ref{equ:DIA_z1.0}) to a value of stellar mass it would seem that $(10^{10}-5\times10^{10})\,{\rm M}_\odot$ stellar mass galaxies are responsible for \DIA$(z=0.0)$, and the $(5\times10^{10}-10^{11})\,{\rm M}_\odot$ range stellar mass galaxies could be responsible for both \DIA$(z=0.5)$ and \DIA$(z=1.0)$. However, it is difficult to disentangle the contributions each mass bin makes to the total intrinsic alignment strength. Thus, we cannot make a definitive determination on this matter. We note that a study by \cite{2015MNRAS.454.2736C} looked at the stellar mass dependence for projected shape alignments in the Horizon-AGN simulation. While they reported an increasing galaxy alignment signal with increasing stellar mass (for stellar masses up to $10^{10.7}\,\rm{M}_{\odot}$), they found no clear identification of which galaxies were responsible for their measured alignment signal.  

\begin{figure*}
    \includegraphics[width=0.99\columnwidth]{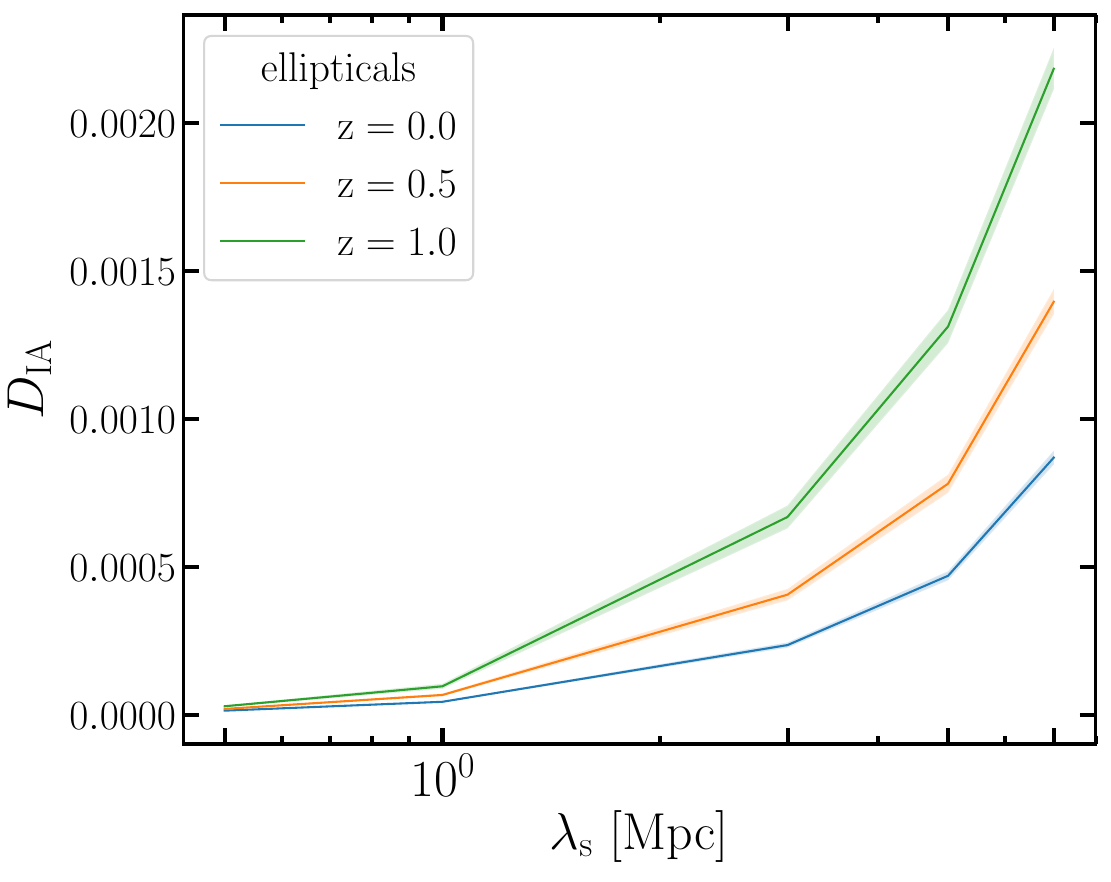}
    \includegraphics[width=0.95\columnwidth]{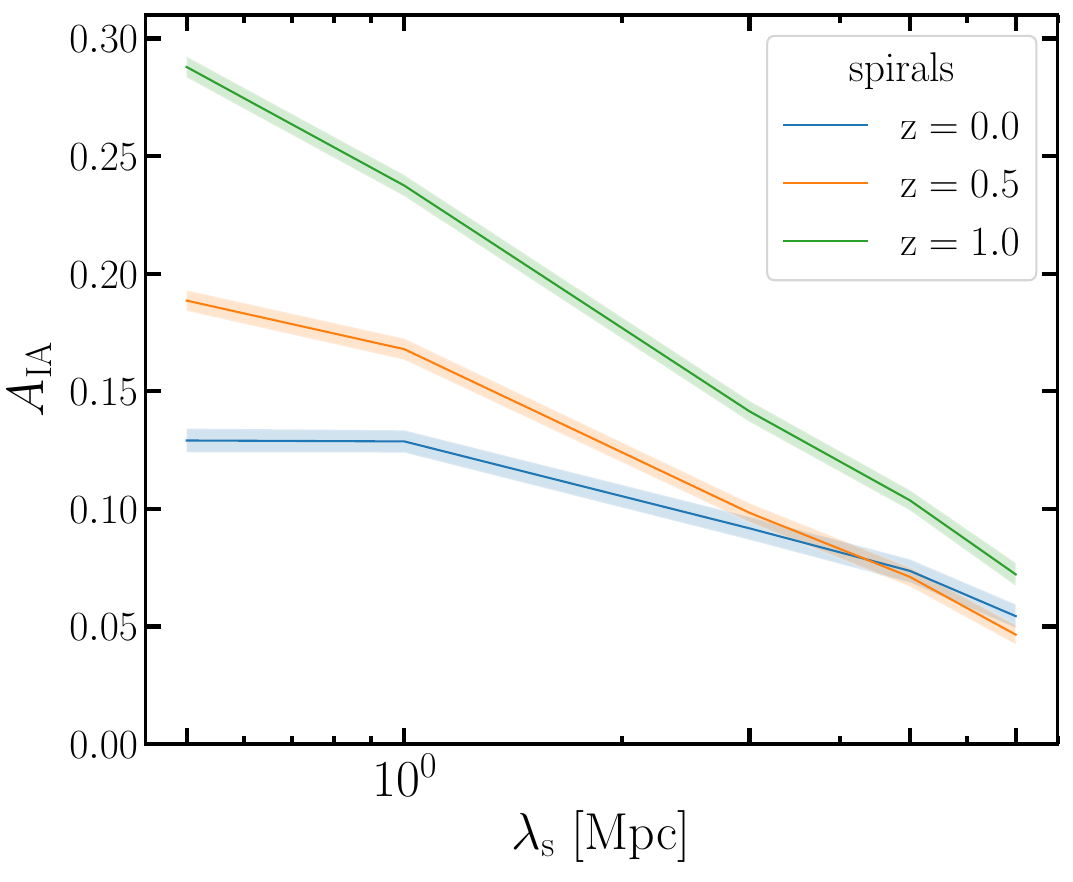}
    \caption{Intrinsic alignment strengths as a function of tidal tensor smoothing scale. Results are shown for galaxies at three different redshifts: $z=0.0$, $z=0.5$, and $z=1.0$. \textit{Left:} The average IA strength for elliptical galaxies, \DIA~, generally shows a greater magnitude with increasing smoothing scale. \textit{Right:} average intrinsic alignment for spiral galaxies, $A_{\mathrm{IA}}$, has the opposite trend with respect to smoothing scale.
    }
    \label{fig:IA_Rs}
\end{figure*}

We further examine the intrinsic alignment strength dependence on the smoothing scale at which the tidal shear tensor is calculated, which can be viewed as a probe of different environments. Fig.~\ref{fig:IA_Rs} shows a clear trend of an increasing \DIA~in the $V$-band as  a function of smoothing scale, $\lambda_s$, for elliptical galaxies. Larger smoothing reduces the amplitude of the tidal field so that the  inferred \DIA values need to increase to compensate while the correlation persists, so  this is not really surprising. Similarly, the right panel of  Fig.~\ref{fig:IA_Rs} shows that the $A_{\rm IA}$ strength decreases with smoothing scale, but this is again largely reflecting the inherent scale-dependence of the definitions of \DIA and $A_{\rm IA}$. It is not obvious, unfortunately, which smoothing scale should be picked for the most meaningful measurement of the large-scale environment that dominates the local tidal field.

\section{Intrinsic Alignment of Halos}
\label{IA_halos}
In this section, we repeat the analysis described in Section~\ref{results:IA_ells}. Specifically we measure the strength of the intrinsic alignment parameter with the linear model, but this time for host dark-matter halos with a minimum halo mass $M_{\rm{halo}}=10^{12}\,$\Msol, of which MTNG740 yielded a sample of $\sim3.3\times10^5$ halos. We calculated \DIA~according to eqn.~(\ref{eq.linear_model}), however, we note that the only difference lies in our definition of  $\epsilon$ for halos; in this section, $\epsilon$ is determined from the positions of the dark-matter particles as opposed to luminosity-weighted positions of stars as is done for galaxies. 

\begin{figure*}
    \centering
    \includegraphics[width=0.95\textwidth]{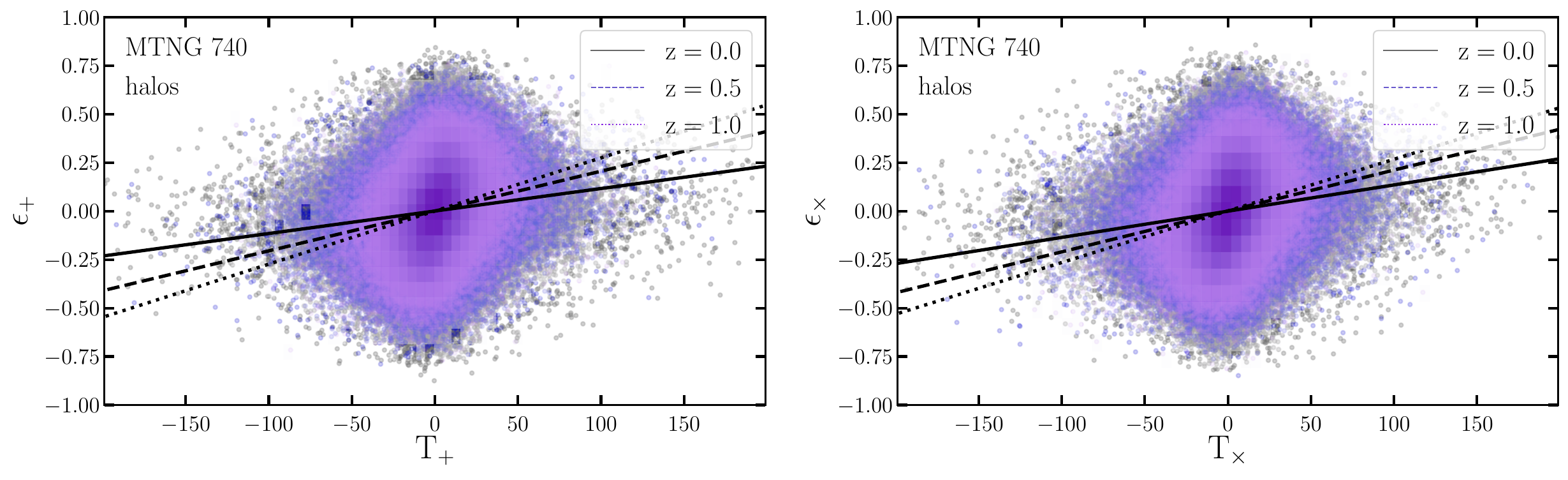}
    \caption{Similar to Fig.~\ref{fig:IA_ellipticals} but for halos. We find significant alignment strengths at three redshifts for the real and imaginary components of ellipticity as a function of the tidal shear tensor calculated at a smoothing scale $\lambda_s = 7\,{\rm Mpc}$.} 
    \label{fig:IA_halos}
\end{figure*}

At $z=0.0$, we measured the tangential component of  ellipticities as a function of tidal shear with a smoothing scale of $\lambda_s = 7\,{\rm  Mpc}$ of dark matter halos in MTNG740, finding an intrinsic alignment parameter value of \DIA$_+ = (1.16\pm0.05)\times10^{-3}$. For the transverse component we obtain \DIA$_{\times} = (1.35\pm0.05)\times10^{-3}$. These values are in close agreement with 15\% difference, and are shown as solid black lines fitted to the light grey data points in Fig.~\ref{fig:IA_halos}. Averaged, they give an intrinsic alignment parameter value of 
\begin{equation}
    D_{{\rm IA}(z=0.0) }= (1.26\pm0.04)\times10^{-3},
\end{equation}
which is $\sim32\,\sigma$ different from null, suggesting significant intrinsic alignment. 

At $z=0.5$, the tangential and transverse components are also in close agreement with only a 2.6\% difference between the two. They are illustrated as dashed lines fitted against blue-grey data points in Fig.~\ref{fig:IA_halos}. We report an average intrinsic alignment parameter value of
\begin{equation}
    D_{{\rm IA}(z=0.5)} = (2.08 \pm 0.07)\times 10^{-3},
\end{equation}
which differs by $\sim30\sigma$ from null, again suggesting considerable alignment. 

At $z=1.0$, the tangential and transverse components of the intrinsic alignment of halos were, again, in very close agreement, about 2.8\% difference between the two. The results are shown in Fig.~\ref{fig:IA_halos} as dotted black lines fitted against purple data points. We obtain an average \DIA~parameter value of
\begin{equation}
    D_{{\rm IA}(z=1.0)} = (2.70\pm0.09)\times10^{-3},
\end{equation}
which is $\sim30\sigma$ away from zero. The value of $D_{{\rm IA}(z=1.0)}$ is about 1.7 times higher than that of $D_{{\rm IA}(z=0.0)}$, suggesting stronger intrinsic alignment of halos at higher redshifts than at the present time, consistent with the trend in galaxy alignments found in the previous sections. 
 
\section{Galaxy-Halo Misalignment}
\label{Misalignmnet}
\begin{figure}
    \centering
    \includegraphics[width=0.95\columnwidth]{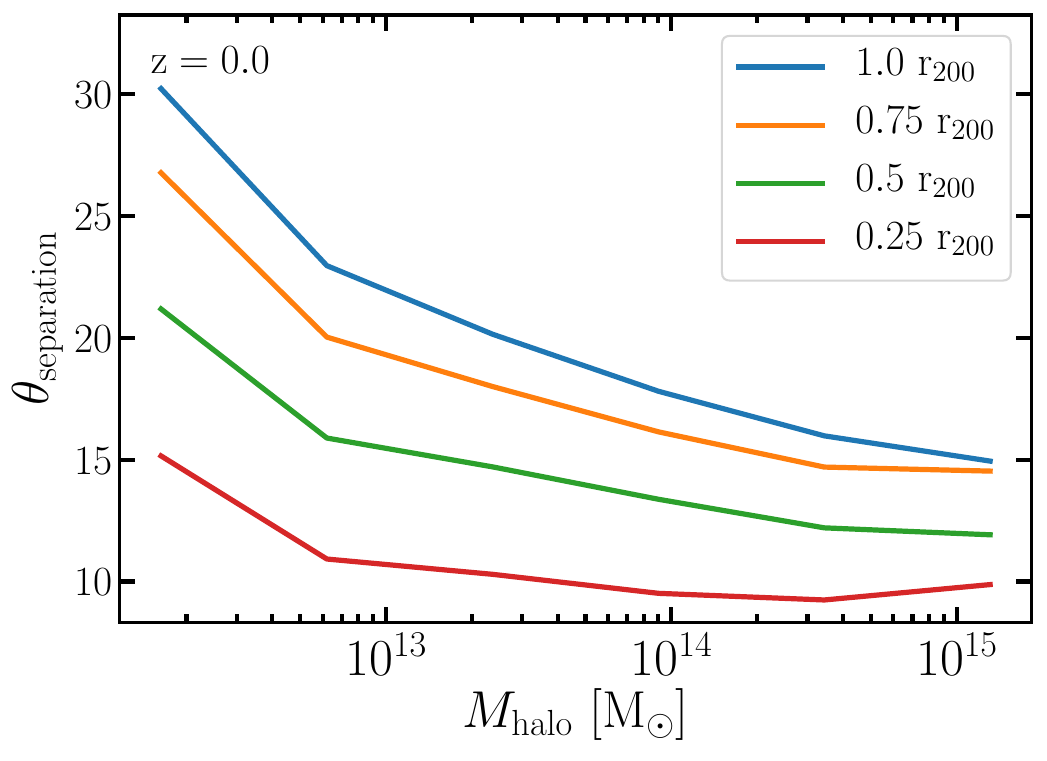}
    \includegraphics[width=0.95\columnwidth]{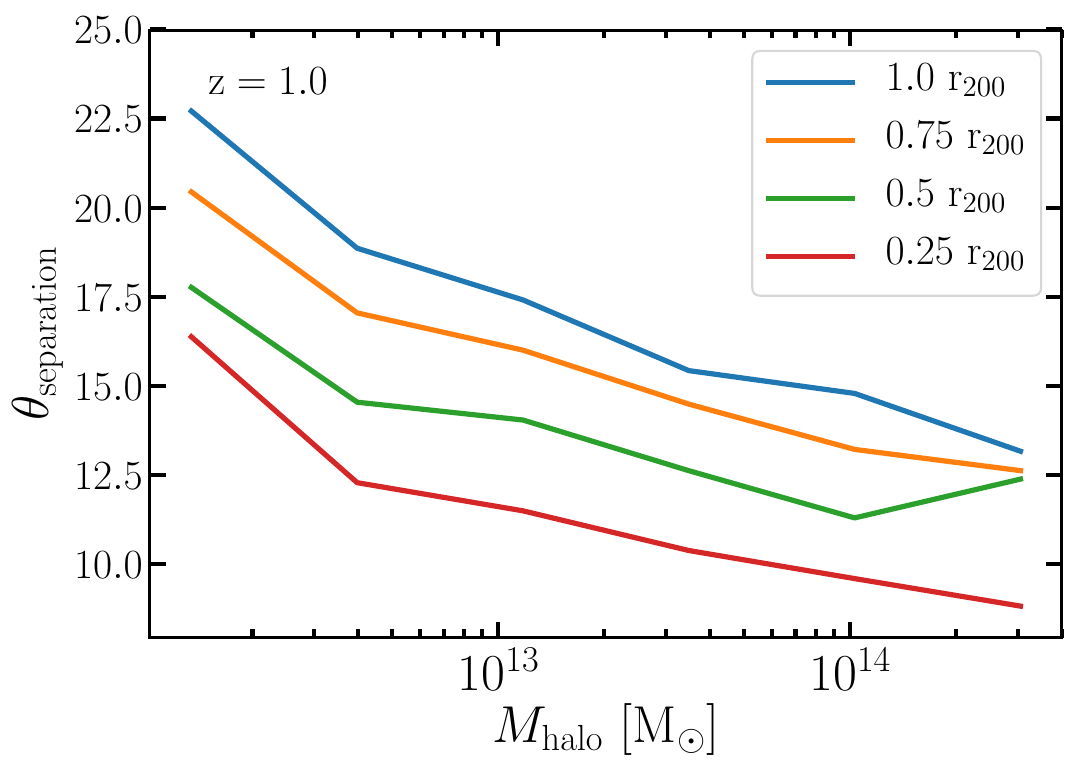}
    \caption{The median misalignment angle between the principal axes of the host halo and its central galaxy as a function of halo mass ($M_{\rm{200m}}$). Results are given for $z=0.0$ in the top panel and $z=1.0$ in the bottom panel. We show this relation for four different fractions of $R_{\rm 200m}$.}
    \label{fig:misalignment}
\end{figure}

In this section we examine the misalignment between host halos with masses $M_{\rm{halo}}>10^{12}\, h^{-1}\,$\Msol~and their central galaxies. To identify the principal axis of each object, we begin by measuring the inertia tensor as:
\begin{equation}
    I_{ij}=\sum^N_{k=1} = m_k\left(||\vec{r}_k||^2\delta_{ij} - x^{(k)}_i x^{(k)}_j\right),
\end{equation}
where $N$ is the total number of particles in the object, $m_k$ is the mass of the $k$-th particle, $\vec{r}_k$ is the 3-dimensional position of the particle, $\delta_{ij}$ is the Kronecker delta, and $x^{(k)}_i$ and $x^{(k)}_j$ are the $i$- and $j$-components of the position of the $k$-th particle, respectively. We compute the inertia tensor of halos using the mass and positions of dark matter particles, while for central galaxies we use stellar particles. We measure the inertia tensors for particles within halo radii $r = R_{{\rm 200m}}$ as defined by the spherical overdensity algorithm applied to each halo, as well as at $0.75\,r$, $0.5\,r$ and $0.25\,r$. The eigenvectors of the inertia tensors are then ordered according to the eigenvalues as $e^1_x, e^2_x, e^3_x$, where $e^1_x$ is the principal axis of the corresponding object (halo or galaxy). Lastly we measure the angle of separation, $\theta$, between $e^1_{\rm{halo}}$ and $e^1_{\rm{galaxy}}$. We note that the principal axis is invariant under rotation of 180 degrees, therefore $\theta$ can only vary between 0-90 degrees.

Fig.~\ref{fig:misalignment} shows the median angle of misalignment, $\theta$, at four fractional halo radii as a function of halo mass, $M_{\rm{200m}}$. We note that there is significant spread from the median, $\sim$20 degrees, at each halo mass bin, although the mean values are only $\sim$5 degrees higher than the reported median values. It would be interesting to investigate the halo-galaxy pairs with misalignment significantly greater than the mean, but we leave that for a future study. We find a general trend of decreasing $\theta$ with increasing halo mass. This is consistent with \citet{2015MNRAS.453..721V} who measured the misalignment between host halos and their central galaxies in the EAGLE and cosmo-OWLS simulations. We used only the dark matter distribution to calculate the inertia tensor of halos, while these authors used the total matter distribution including dark matter, stars and gas. We believe the significantly higher number of dark matter particles should dominate sufficiently such that including the baryon particles would not make a statistical difference in our measurement, however. 

\section{Summary and Conclusions}
\label{conclusions}
This paper is part of a series of introductory papers of the MillenniumTNG (MTNG) project, a new simulation series which combines the galaxy formation model of IllustrisTNG \citep{2017MNRAS.465.3291W,2018MNRAS.473.4077P} with the large cosmological volume of the Millennium simulation \citep{Springel2005}. In this work we employ the flagship full physics simulation, MTNG740, with a comoving volume of (740 Mpc)$^3$, baryonic mass resolution of $3.1\times10^7$ \Msol~and dark-matter mass resolution of $1.7\times10^8$ \Msol~to examine the projected correlation of intrinsic galaxy shapes, as well as the intrinsic alignments of galaxies and dark matter halos against their local tidal field at redshifts $z=0.0$, $z=0.5$ and $z=1.0$.  Thanks to our large simulation volume, we reach a statistically more robust population of galaxies that is less affected by cosmic variance than previous work. For this reason we need not perform random rotations of our galaxies to mitigate cosmic variance effects such as in Z22 for TNG100. 

We perform three different measurements of the projected correlation of intrinsic galaxy shear against the background large-scale distribution: intrinsic galaxy shear with galaxy positions as tracers, $w_{\rm p}^{+g}$, direct correlations with intrinsic shear of other galaxies $w_{\rm p}^{++}$, and correlations with the total matter field using dark matter and baryonic particles/cells in MTNG740, $w_{\rm p}^{+m}$. These projected correlations are measured for our entire galaxy population as well as for subsamples of elliptical and spiral galaxies. 

We further measure the intrinsic alignment of galaxies with their local tidal field tensor on a 7~Mpc smoothing scale against two different models based on galaxy morphology. We define the shape of elliptical galaxies by the luminosity-weighted distribution of stars and measure the intrinsic alignment strength, \DIA, under a linear model as given by equation~(\ref{eq.linear_model}). For spiral galaxies, we first define their shape based on the angular momentum of the stellar component in the disk, and then measure the intrinsic alignment parameter, $A_{\rm{IA}}$, under the quadratic model as defined by equation~(\ref{eq.quadratic}). Mass and environmental dependence of \DIA~and $A_{\rm{IA}}$ were also examined in this study.

Considering that the alignment of elliptical galaxies is thought to coincide with the orientation of their host halos, we also measure the intrinsic alignment of dark-matter halos against the value of the tidal shear tensor at a 7 Mpc smoothing scale. We use the linear alignment model to measure \DIA~for halos, where we define the shape of the halo in the same manner as for elliptical galaxies but using the unweighted positions of the dark-matter particles.

In a final analysis, we examine the misalignment angle between host halos with masses $M_{\rm{halo}}>10^{12} h^{-1}$ \Msol~and their central galaxies. We measure the principal axis of halos based on their dark-matter particle distribution, and for galaxies we determine the principal axis based on the stellar distribution. The misalignment angle is defined as the angle of separation between their respective principal axes. We measure the misalignment at four radii within the host halo, and furthermore look at its dependence on total mass of the host halo. 
\ \\

\noindent
We obtain the following primary results: \vspace*{-0.2cm}
\begin{enumerate}
\item We find a strong, high signal-to-noise intrinsic alignment signal in the $w_{\rm p}^{+g}$ correlation function, whose amplitude and shape is in good agreement with observational intrinsic alignment measurements for the LOWZ sample of the SDSS-III BOSS survey obtained by \cite{2015MNRAS.450.2195S}. Our results are also in reasonable agreement with the  intrinsic alignment measures of galaxies from hydrodynamical simulations presented in \cite{2021MNRAS.508..637S}, which, however, exhibit much larger statistical errors due to smaller simulation volumes. We further note a strong redshift dependence in $w_{\rm p}^{+g}$, as well as a dependence on galaxy morphology (Fig.~\ref{fig:wp_plot}).

\item Shear-shear correlations, $w_{\rm p}^{++}$, are significant for elliptical galaxies in both the one- and two-halo regimes out to large distances, although they are much stronger in the former than in the latter. At face value, there is a clear redshift dependence of $w_{\rm p}^{++}$ for our elliptical and all galaxy samples. This is largely driven by the evolving galaxy bias that similarly manifests itself in the evolution of $w_{\rm p}^{+g}$, whereas the effective alignment signal itself, $w_{\rm p}^{++}/w_{\rm p}^{+g}$, is almost redshift-independent. In contrast, $w_{\rm p}^{++}$ is consistent with zero for spiral galaxies in the two-halo regime, with only a marginal detection of a non-zero amplitude for $r_{\rm p}\lesssim 1\, h^{-1}{\rm Mpc}$.

\item Significant intrinsic alignments are also found in $w_{\rm p}^{+m}$ for both elliptical and spiral galaxies. However, we note an opposite redshift dependence compared to that of $w_{\rm p}^{+g}$: $w_{\rm p}^{+m}$ decreases with increasing redshift. This gives evidence that the growth of amplitude with redshift in $w_{\rm p}^{+g}$ can be attributed to an evolving galaxy bias, as we have explicitly verified (Fig.~\ref{fig:wp_bias_plot}). 

\item We find a significant intrinsic alignment strength for elliptical galaxies (under a linear model) at redshifts $0.0 \leq z \leq 1.0$, with increasing alignment strength at higher redshifts, consistent with Z22 (Fig.~\ref{fig:IA_ellipticals}). We also find agreement between the tangential and cross components of the intrinsic alignment, and measure average parameter values of \DIA$_{(z=0.0)}=(8.79\pm0.20)\times10^{-4}$ and \DIA$_{(z=1.0)}=(2.21\pm0.06)\times10^{-3}$, respectively.

\item The intrinsic alignment of spiral galaxies (under a quadratic model) is significantly weaker  than ellipticals under a linear model, consistent with Z22.

\item The value of the intrinsic alignment parameter, \DIA, depends on galaxy and environmental properties. \DIA~increases when the smoothing scale of the tidal shear tensor, $\lambda_s$, is increased (Fig.~\ref{fig:IA_Rs}), because the smoothing reduces the amplitude of the tidal field so that \DIA becomes larger to compensate while the correlation is still intact. We also find that \DIA~generally increases for galaxies with higher stellar mass, consistent with Z22, \cite{2015MNRAS.454.2736C}, \cite{2015MNRAS.448.3522T} and \cite{2015MNRAS.454.3328V}. 

\item We find a significant intrinsic alignment strength for dark-matter halos with a clear redshift dependence (Fig.~\ref{fig:IA_halos}). At a tidal shear tensor smoothing scale of 7 Mpc we measure \DIA$_{(z=0.0)} = (1.26\pm0.04)\times10^{-3}$ and \DIA$_{(z=1.0)} = (2.70\pm0.09)\times10^{-3}$, respectively.

\item We report small to moderate misalignments between host dark-matter halos and their central galaxies in MTNG740, and find that the misalignment angles, $\theta$, are generally similar at  $z=0.0$ and $z=1.0$. We further detect mass and environmental dependencies in $\theta$, which decrease as halo mass increases, and decrease with decreasing halo radius.

\end{enumerate}

In summary, we measured the intrinsic alignments, IA, of galaxies as a function of morphology and redshift  and found strong IA signals for both elliptical and spiral galaxies by way of the projected shear correlations of galaxy shapes, $w_{\rm p}^{+g}$, $w_{\rm p}^{+m}$, and $w_{\rm p}^{++}$, as well as by measuring the alignment strengths of elliptical galaxies against the local tidal field under a linear model. The agreement with observational measurements and with previous studies of IA in hydrodynamical simulations is encouraging and suggests that MTNG740 can be a powerful tool for making realistic predictions of the effects of IA on weak lensing statistics. Additionally, the fact that we found negligible IA signal for spiral galaxies under a quadratic model supports recent literature arguing for one intrinsic alignment model for all galaxy types to be used in weak lensing pipelines.

While the IA measurements in this study are in good qualitative agreement with previous literature and reproduce currently available observational results, they also exhibit much smaller statistical errors than obtained previously, thanks to the unprecedented statistical power of MTNG740 compared with any previous hydrodynamical simulation used in this context. This offers the prospect of using the MTNG simulations in future work for detailed  forward modelling of lensing observations  including IA effects, as well as to gain better understanding of the physical origin of these effects.

\section*{Acknowledgments}
We wish to thank Sergio Contreras, N. Elisa Chisari and Claire Lamman for helpful comments and feedback in this work. 
CH-A acknowledges support from the Excellence Cluster ORIGINS which is funded by the Deutsche Forschungsgemeinschaft (DFG, German Research Foundation) under Germany's Excellence Strategy -- EXC-2094 -- 390783311. 
VS and LH acknowledge support by the Simons Collaboration on ``Learning the Universe''. 
LH is supported by NSF grant AST-1815978.  
SB is supported by the UK Research and Innovation (UKRI) Future Leaders Fellowship [grant number MR/V023381/1]. 
The authors gratefully acknowledge the Gauss Centre for Supercomputing (GCS) for providing computing time on the GCS Supercomputer SuperMUC-NG at the Leibniz Supercomputing Centre (LRZ) in Garching, Germany, under project pn34mo. 
This work used the DiRAC@Durham facility managed by the Institute for Computational Cosmology on behalf of the STFC DiRAC HPC Facility, with equipment funded by BEIS capital funding via STFC capital grants ST/K00042X/1, ST/P002293/1, ST/R002371/1 and ST/S002502/1, Durham University and STFC operations grant ST/R000832/1.

\section*{Data Availability}
The MillenniumTNG simulations will be made publicly available by the end of 2023. Figures from this work may be made available upon request. 

\bibliographystyle{mnras}
\bibliography{main}

\label{lastpage}
\end{document}